\documentclass[letterpaper,twocolumn,pra,aps,showpacs,floatfix]{revtex4-1}
\usepackage{amsmath,amssymb,graphicx}
\usepackage{CJK}
\usepackage[bookmarks=true,colorlinks,linkcolor=red,urlcolor=blue,citecolor=blue]{hyperref}
\usepackage[usenames]{color}
\usepackage{float}
\usepackage{subfigure}
\usepackage{bbm}

\newcommand{\Da}{{\dagger}}
\newcommand{\eat}[1]{}

\newcommand{\bmr}{\boldsymbol{ \mathsf{r} } }
\newcommand{\bsr}{\boldsymbol{ r } }
\newcommand{\ba}{\boldsymbol{a} }
\newcommand{\bR}{\boldsymbol{R} }
\newcommand{\br}{\boldsymbol{r} }
\newcommand{\sfr}{\boldsymbol{\mathsf{r}}}
\newcommand{\bJ}{\boldsymbol{J}}
\newcommand{\bj}{\boldsymbol{j}}
\newcommand{\mxyb}{M_{x \bar{y}}}
\newcommand{\bb}{\boldsymbol{b}}
\newcommand{\bk}{\boldsymbol{k}}
\newcommand{\bq}{\boldsymbol{q}}

\begin{document}
\title{Quantum spin ices and topological phases from dipolar-octupolar doublets on the pyrochlore lattice}
\author{Yi-Ping Huang}
\author{Gang Chen}
\altaffiliation[Current address: ]{Department of Physics, University of Toronto, Toronto, Ontario M5S 1A7, Canada}
\author{Michael Hermele}
\affiliation{Department of Physics, 390 UCB, University of Colorado, Boulder,
Colorado 80309, USA}
\date{\today}
\begin{abstract}
We consider a class of $d$- and $f$-electron systems in which dipolar-octupolar Kramers doublets arise on the sites of the pyrochlore lattice.  For such doublets, two components of the pseudospin transform like a magnetic dipole, while the other transforms like a component of the magnetic octupole tensor.  Based on a symmetry analysis, we construct and study models of dipolar-octupolar doublets in itinerant and localized limits.  In both limits, the resulting models are of surprisingly simple form.  In the itinerant limit, we find topological insulating behavior.  In the localized limit, the most general nearest-neighbor spin model is the XYZ model.  We show that this XYZ model exhibits \emph{two distinct} quantum spin ice (QSI) phases, that we dub dipolar QSI, and octupolar QSI.  We conclude with a discussion of potential relevance to real material systems.
\end{abstract}

\date{\today}


\maketitle

Finding new phases of matter is a problem of fundamental importance in condensed matter
physics.  This search motivates in part the exploration of new classes of materials, where novel
parameter regimes can lead to phases not realized elsewhere, and other new phenomena. 
Recently, there has been intense interest in materials combining strong spin-orbit coupling 
(SOC) with substantial electron correlation, especially in compounds with heavy elements 
\cite{Witczak-Krempa2013}.  SOC entangles the spin and orbital degrees of freedom, and 
microscopic models including SOC have in many cases not yet been constructed and studied.  
Spin-orbital entanglement can lead to rather complicated models, but this need not always be 
the case.  

In this letter, we study a class of systems where strong SOC leads to surprisingly
 simple microscopic models that -- in different limits -- naturally realize not only a topological
 band insulator, but also \emph{two distinct} quantum spin ice (QSI) phases.  One of these is the familiar QSI phase \cite{Hermele2004,Gingras2013}, here dubbed \emph{dipolar} QSI (dQSI), while the other is a novel \emph{octupolar} QSI (oQSI).   dQSI and oQSI  are two distinct symmetry enriched ${\rm U}(1)$ quantum spin liquids, with  space group symmetry playing the crucial role.
  
Much of the recent activity in strong-SOC systems has focused on $5d$ iridates and $4f$ pyrochlores.
Various novel models and phases have been predicted for iridates with pyrochlore
\cite{Pesin2010,Wan2011,Kargarian2011,Chen2012,Witczak2012,Witczak2012b}, 
hyperkagome\cite{Chen2008,Zhou2008,Lawler2008,Lawler20082,Bergholtz2010,Chen2013}, 
honeycomb\cite{Jackeli2010} and hyperhoneycomb lattices\cite{Ashvin2013,Kim2013}, while the
 dQSI phase has been predicted in $4f$ pyrochlores\cite{Molavian2007,Onoda2011,Savary2012,Lee2012,Applegate2012}. 
In many of these systems, SOC and other
 interactions lead to Kramers doublets on the $d$ or $f$ ions, which in turn are the building
 blocks for minimal effective models to capture the low-energy physics.  Any Kramers doublet 
is associated with a time-reversal odd pseudospin operator $\tau^{\mu}$ ($\mu = x,y,z$), but not all Kramers doublets transform identically under space
 group symmetry \cite{AbragamBleaney201207}.  The most familiar possibility, which holds in
 the above recently studied $4f$ and $5d$ systems, is that, just like a true spin-1/2 moment, $\tau^{\mu}$ transforms as a magnetic dipole (\emph{i.e.} as a pseudovector) under space group
 operations. 

In this letter, focusing on the pyrochlore lattice of corner-sharing tetrahedra, we consider a class 
of systems with Kramers doublets arising from $d$ or $f$ ions, where (in suitable local coordinates 
discussed below) $\tau^z$ and $\tau^x$ both transform like the $z$-component of a magnetic dipole,
while $\tau^y$ transforms as a component of the magnetic octupole tensor.  
Models of such dipolar-octupolar (DO) doublets have striking properties in both 
weak and strong correlation limits.  We note that a similar type of Kramers doublet has been considered on other lattices\cite{Jackeli2009, Chen2010}.

More specifically we consider both A$_2$B$_2$O$_7$ pyrochlores and AB$_2$O$_4$ spinels, 
where the pyrochlore A-site, and B-sites in both families, form a pyrochlore lattice. 
We consider two principal situations:  (1) In both pyrochlores and spinels, 
B is a transition metal in $d^1$ or $d^3$ electron configuration and A is non-magnetic.   
(2) In pyrochlores, A is a trivalent rare earth with a partially filled $4f$ shell, and B is non-magnetic.
Both cases can lead to effective models of DO doublets on the pyrochlore lattice.

\emph{Case $(1)$.}  The magnetic ions reside at the center of a trigonally-distorted oxygen octahedron; 
the single-ion physics has been treated \emph{e.g.} in \cite{AbragamBleaney201207}.  
Due to the cubic crystal field only the $t_{2g}$ manifold is relevant.  Projection ${\cal P}$ of orbital angular 
momentum $\boldsymbol{L}$ into the $t_{2g}$ manifold is ${\cal P} \boldsymbol{L} {\cal P} = - \boldsymbol{\ell}$, 
where the $\ell^{\mu}$ are spin-1 matrices.  The single-site Hamiltonian within the $t_{2g}$ manifold is
\begin{equation}
H = -\lambda \,\boldsymbol{\ell} \cdot \boldsymbol{S} + H_{{\rm tri}} + H_{{\rm int}} \text{,}
\end{equation}
with $\lambda$ the strength of SOC and $\boldsymbol{S}$ the spin operator.  $H_{{\rm tri}}= \Delta_3 (\ell^{z_i})^2  $ 
is the trigonal crystal field allowed by $D_{3d}$ site symmetry. The $z_i$-axis is 
the local $C_3$ axis ($i = 1,\dots4$ is the sublattice index), and $x_i, y_i$-axes are specified in the supplementary material\cite{Note1}.
The interaction $H_{{\rm int}}$ is of Kanamori form, and is treated in the atomic limit where it 
is characterized by Hubbard interaction $U$ and Hund's coupling $J_H$ \cite{Note1}.

Defining an effective total angular momentum $\boldsymbol{j}_{{\rm eff}}=\boldsymbol{\ell}+\boldsymbol{S}$, 
SOC alone splits the $t_{2g}$ manifold into an upper doublet ($j_{{\rm eff}} = 1/2$) and lower quadruplet 
($j_{{\rm eff}} = 3/2$).  Effective models of $j_{{\rm eff}} = 1/2$ doublets are relevant for $5d^5$ iridates \cite{BJKim2008,BJKim2009} and have 
received significant attention\cite{Chen2008,Pesin2010,Jackeli2010,Witczak2012,Ashvin2013,Kim2013}. 
While the $j_{{\rm eff}} = 1/2$ doublet is dipolar, it does not obey a na\"{\i}ve Heisenberg exchange model 
due to strong SOC \cite{Chen2010,Chen2011}.

\begin{figure}[t]
	\centering
	\includegraphics[width=\columnwidth]{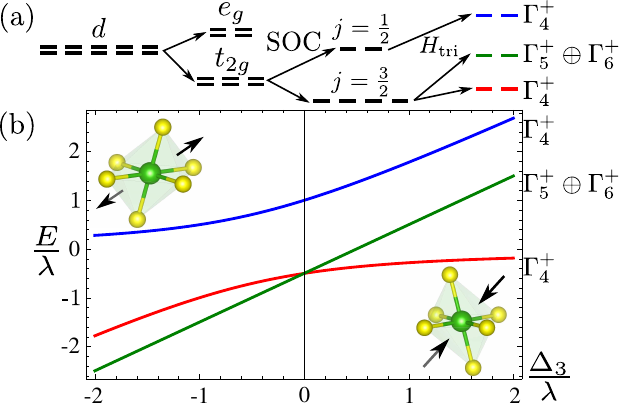}
	\caption{ (Color online.)
(a) The evolution of $d$ electron states under cubic crystal field, SOC and trigonal distortion.
(b) The energies for the three local doublets under different trigonal distortions. 
Compression (elongation) along the $C_3$ axis corresponds to $\Delta_3>0$ ($\Delta_3<0$).}
	\label{fig:fig1}
\end{figure}

The trigonal crystal field $H_{{\rm tri}}$ splits the 
quadruplet into two Kramers doublets, for a total of three doublets. If $\Delta_3 > 0$,
the lower and upper doublets are dipolar and transform as the $\Gamma^+_4$ irreducible 
representation of the $D_{3d}$ double group \cite{KosterBook}.  The middle doublet is a DO doublet; 
it has $j^{z_i}_{{\rm eff}} = \pm 3/2$, and transforms as $\Gamma^+_5 \oplus \Gamma^+_6$
(Fig.~\ref{fig:fig1}).  
The doublet is half-filled for $d^3$ electron configuration, or (if $\Delta_3 < 0$) for $d^1$ configuration.

While Hubbard interaction does not affect the single-site energy spectrum for a fixed number of electrons, 
Hund's coupling plays an important role.  When
${\Delta_3} > 0$, we find  the $d^3$ ground state multiplet remains a DO doublet even for large $J_H$ \cite{Note1}.  However, as $J_H$ increases, the energy gap between the ground state and the dipolar doublet first excited state decreases, vanishing in the limit of large $J_H$ where we recover a spin-3/2 moment.  The splitting between the ground and first excited doublets is substantial only when $J_H \lesssim \lambda$, and increases with $\Delta_3 / \lambda$ \cite{Note1}.  Hund's coupling has no effect for $d^1$ configuration.

\emph{Case $(2)$.}  Here A is a trivalent rare earth, where the ground state has angular momentum $J$.  The $D_{3d}$-symmetric crystal field Hamiltonian is $H_{{\rm cf}} = 3 B^0_2 (J^z)^2 + \cdots$ \cite{Gardner2010}.  If $J = 9/2$ or $15/2$, and $B^0_2 < 0$ and dominates the other crystal field terms, then the ground state is a DO doublet with $J^z = \pm J$, transforming as $\Gamma^+_5 \oplus \Gamma^+_6$ under $D_{3d}$ site symmetry.  The DO doublet nature of the ground state is robust even when the other crystal field terms are appreciable, as long as the ground state is adiabatically connected to the $J^z = \pm J$ doublet.  Among the lanthanides, only Nd$^{3+}$, Dy$^{3+}$ and Er$^{3+}$ have the required values of $J$. Of these, $B^0_2 < 0$ only for Nd$^{3+}$ and Dy$^{3+}$ \cite{Gardner2010}. Indeed, the crystal field ground state of  Nd$^{3+}$ in Nd$_2$Ir$_2$O$_7$ is a DO doublet \cite{Watahiki2011}, and a DO doublet ground state is predicted for Dy$^{3+}$ in Dy$_2$Ti$_2$O$_7$\cite{Bertin2012}.

The action of $Fd\bar{3}m$ space group symmetry on DO doublets is given in the supplementary material \cite{Note1}. The $D_{3d}$ site symmetry is generated by a 3-fold rotation $C_3$, a mirror plane $M$, and inversion ${\cal I}$, with:  $C_3 : \tau^{\mu} \to \tau^{\mu}$, $M : \tau^{x, z} \to - \tau^{x, z}$, $M : \tau^{y} \to \tau^y$, and ${\cal I} : \tau^{\mu} \to \tau^{\mu}$.  These transformations are not those of a pseudovector, and imply that $\tau^{x,z}$ transform like the  $z_i$-component of a magnetic dipole, while $\tau^y$ transforms like a component of the magnetic octupole tensor \cite{Note1}.

We now proceed to construct effective models 
using a single DO doublet on each pyrochlore lattice site as the basic building block.  
We assume throughout that higher-energy on-site degrees of freedom can be ignored.  
Even when this is not quantitatively accurate, our models may still be valid as minimal low-energy effective models.  

We consider limits of itinerant and localized electrons, constructing tight-binding and spin Hamiltonians, 
respectively, in the two limits. The Hamiltonian contains all electron hopping terms (itinerant limit) or spin 
exchange terms (localized limit) allowed by time reversal and $Fd\bar{3}m$ space group symmetry, 
up to a given spatial range. 
We note that  tight-binding and exchange models of dipolar $\Gamma_4^+$ doublets have been extensively 
studied in the context of iridate and rare-earth pyrochlores\cite{Curnoe2008,Pesin2010,Kurita2011,Witczak-Krempa2012,Witczak2012b,Lee2012,Savary2012}.

In the itinerant limit we ignore electron interactions, and the general form of the model is 
\begin{equation}
H_{TB} = \sum_{(\br, \br')} \big[ c^\dagger_{\br} T_{\br \br'} c^{\vphantom\dagger}_{\br} + h.c. \big] \text{.}
\end{equation}
Here, $\br$ labels pyrochlore lattice sites, the sum is over all pairs of sites, and
$c^T_{\br} = (c_{\br +}, c_{\br -} )$.  
$T^{\vphantom\dagger}_{\br \br'} = T^\dagger_{\br' \br}$ is a $2 \times 2$ matrix describing tunneling between sites $\br$ and $\br'$.  
The operator $c^\dagger_{\br \pm}$ creates an electron at site $\br$ with $j^{z_i}_{{\rm eff}} = \pm 3/2$ in case (1), or $J^{z_i} = \pm J$ in case (2).  Pseudospin operators are $\tau^{\mu}_{\br} = (1/2) c^\dagger_{\br} \sigma^{\mu} c^{\vphantom\dagger}_{\br}$, 
where $\sigma^{\mu}$ are the Pauli matrices. Time reversal symmetry implies 
$T_{\br \br'} = t_{\br \br'}^0 + i t_{\br \br'}^{\mu} \sigma^{\mu}$.

For nearest-neighbor sites, the hopping matrix $T_{\br \br'}$ has a remarkably simple form.  
Choosing an appropriate orientation of bonds \cite{Note1}, we find 
$T_{\br \br'} = i [ t^1_{nn} \sigma^1 + t^3_{nn} \sigma^3 ]$, \emph{taking the same 
form for all nearest-neighbor bonds}.  A global rotation about the $y$-axis in 
pseudospin space can thus eliminate $t^1_{nn}$, leading to $\tilde{T}_{\br, \br'} = i \tilde{t}^3_{nn} \sigma^3$, 
where the tilde indicates we are working in the transformed basis.  
The nearest-neighbor model thus has a ${\rm U}(1)$ spin symmetry, and the purely imaginary 
(spin-dependent) hopping is similar to models considered in \cite{Burnell2009}.  
A highly unstable Fermi surface coincides with a surface of intersection between two bands\cite{Note1}.

\begin{figure}[t]
	\centering
	\includegraphics[width=\columnwidth]{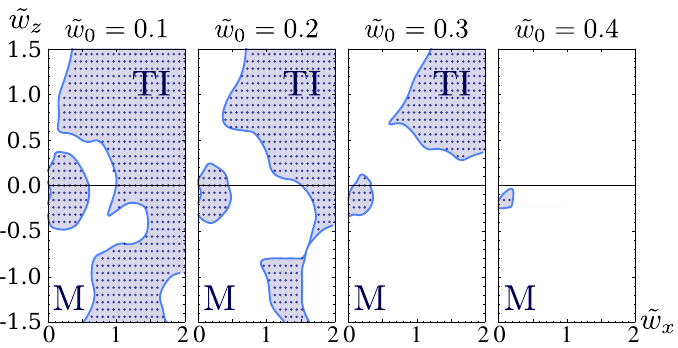}
         \caption{Phase diagram of the tight-binding model wtih first- and second-neighbor hopping, as a function of $(\tilde{w}_0, \tilde{w}_x, \tilde{w}_z)$, setting $\tilde{t}^3_{nn}=1$.  Very small fourth-neighbor hopping is included to remove unstable band touchings at the W-point.  Metallic (M) and strong topological insulator (TI) phases are found. The phase diagram is symmetric under $\tilde{w}_x \to - \tilde{w}_x $ and $\tilde{w}_0 \to -\tilde{w}_0$.}
         \label{fig:fig2}
\end{figure}

Evidently the nearest-neighbor tight-binding model is highly fine-tuned, 
so we also include second-neighbor hopping, which is  specified 
by parameters $(\tilde{w}_0, \tilde{w}_x, \tilde{w}_z)$ \cite{Note1}. Second-neighbor hopping breaks the  ${\rm U}(1)$ 
spin symmetry and gaps out most of the nearest-neighbor Fermi surface.  One finds either a metallic state, or a semi-metal with isolated band touchings occurring at the W points (see  Fig.~\ref{fig:fig2}).  
These W-point touchings are in fact unstable and are gapped out by fourth-neighbor hopping, leading to a strong topological band insulator \cite{Note1}.

We now consider the large-$U$ limit of localized electrons, where the degrees of freedom 
are the pseudospin-$1/2$ moments  $\tau^{\mu}_{\br}$.  We find that the most general symmetry
allowed nearest-neighbor exchange is 
$H_{ex} = \sum_{\langle \br \br' \rangle} [
J_x \tau^x_{\br} \tau^x_{\br'} + J_y \tau^y_{\br} \tau^y_{\br'} + J_z \tau^z_{\br} \tau^z_{\br'}
+ J_{xz} (\tau^x_{\br} \tau^z_{\br'} + \tau^z_{\br} \tau^x_{\br'} )]$, where the sum is over nearest-neighbor bonds.
Quite remarkably, the exchange is identical in form on every bond. 
Similar to the itinerant limit,  the $J_{xz}$ term can be eliminated by a global pseudospin rotation\cite{Note1}.
After this transformation, the exchange is of the remarkably simple XYZ form:
\begin{eqnarray}
	H_{\text{XYZ}} = \sum_{\langle \br \br' \rangle}
	\tilde{J}_{x} \tilde{\tau}_{\br}^x \tilde{\tau}_{\br'}^x 
     + 	\tilde{J}_{y} \tilde{\tau}_{\br}^y \tilde{\tau}_{\br'}^y
     + \tilde{J}_{z} \tilde{\tau}_{\br}^z \tilde{\tau}_{\br'}^z \text{.} 
	\label{ex_model}
\end{eqnarray}
This result should be contrasted with the case of dipolar doublets on the pyrochlore lattice, where the form of nearest-neighbor exchange varies according to the orientation of each bond \cite{Onoda2011}. 

Beyond simplicity of form, this pyrochlore XYZ model supports two distinct QSI phases.  
To see this, we first review the XXZ model ($\tilde{J}_\perp \equiv \tilde{J}_x = \tilde{J}_y$), where  QSI  was identified in a study of the regime $\tilde{J}_z > 0$, $\tilde{J}_z \gg | \tilde{J}_\perp |$ \cite{Hermele2004}.  For simplicity we concentrate on $\tilde{J}_\perp < 0$, where quantum Monte Carlo  \cite{Banerjee2008} found QSI for $ | \tilde{J}_\perp | / \tilde{J}_z < c$, with $c \approx 0.1$.  When $ | \tilde{J}_\perp | / \tilde{J}_z > c$, magnetic order is present. 
It is important to note that QSI is robust to arbitrary symmetry breaking perturbations, and thus survives away from the XXZ line.

The physics of QSI  can be understood by mapping to a compact ${\rm U}(1)$ gauge theory, which is exact for large $\tilde{J}_z$ \cite{Hermele2004}.  The centers of pyrochlore lattice tetrahedra $\bmr$ form a diamond lattice, and each pyrochlore site $\br$ corresponds to a unique  nearest-neighbor diamond link $(\bmr, \bmr')$.  We introduce lattice vector fields $ E_{\bmr \bmr'} = \tilde{\tau}^z_{\br}$ and $e^{i A_{\bmr \bmr'} } = \tilde{\tau}^x_{\br} + i \tilde{\tau}^y_{\br}$,
where $\bmr$ ($\bmr'$) lies in the diamond A (B) sublattice, and $E_{\bmr \bmr'} = - E_{\bmr' \bmr}$, $A_{\bmr \bmr'} = - A_{\bmr' \bmr}$.  $E$ ($A$) can be interpreted as the electric field (vector potential) of a compact ${\rm U}(1)$ lattice gauge theory, of which QSI is the deconfined phase, supporting a gapless photon, and gapped electric charge and magnetic monopole excitations.

So far we have been describing  dQSI, so named because the electric field $E_{\bmr \bmr'} = \tilde{\tau}^z_{\br}$ is a magnetic dipole. In the low-energy continuum theory, the electric field is odd under time reversal and transforms under the $\Gamma^+_4$ (pseudovector) representation of the $O_h$ point group. [The magnetic field is time reversal even, and transforms under the $\Gamma^-_4$ (vector) representation.]  The same dQSI phase occurs for large $\tilde{J}_x > 0$ ($\tilde{J}_{y,z} < 0$ for simplicity), where $E_{\bmr \bmr'} = \tilde{\tau}^x_{\br}$, which transforms identically to $\tilde{\tau}^z_{\br}$ under symmetry.

The novel oQSI phase arises for $\tilde{J}_y > 0$ large ($\tilde{J}_{x,z} < 0$ for simplicity), so that $E_{\bmr \bmr'} = \tilde{\tau}^y_{\br}$.  In this case the electric field is purely octupolar.  In the continuum theory, the electric field is still time reversal odd, but transforms under the $\Gamma^+_5$ representation of $O_h$ (neither vector nor pseudovector). The magnetic field transforms as $\Gamma^-_5$.

oQSI and dQSI  are thus distinguished by the action of space group symmetry on electric and magnetic fields, and can be viewed as distinct symmetry enriched ${\rm U}(1)$ quantum spin liquids.  This means that dQSI and oQSI are distinct phases in the presence of space group symmetry, but weak space-group-breaking perturbations take dQSI and oQSI into the same ${\rm U}(1)$ quantum spin liquid phase (which is robust to arbitrary weak perturbations regardless of symmetry).  In terms of physical properties, dQSI and oQSI both have a $T^3$ contribution to specific heat from gapless photons; in $f$-electron realizations, this is expected to be about 1000 times the phonon contribution \cite{Savary2012}.  Dipolar spin correlations, as measured \emph{e.g.} by neutron scattering, will, however, be quite different, as illustrated by the fact that, neglecting effects of long-range dipolar interaction, equal-time dipolar correlations fall off as $1/r^4$ in dQSI \cite{Hermele2004}, but as $1/r^8$ in oQSI \cite{Note1}.  In future work, it would be interesting to compare the dynamic spin structure factor in dQSI and oQSI.  Neutron scattering signatures of dQSI have been discussed in \cite{Savary2012}.

\begin{figure}[t]
	\centering
         \includegraphics[width=\columnwidth]{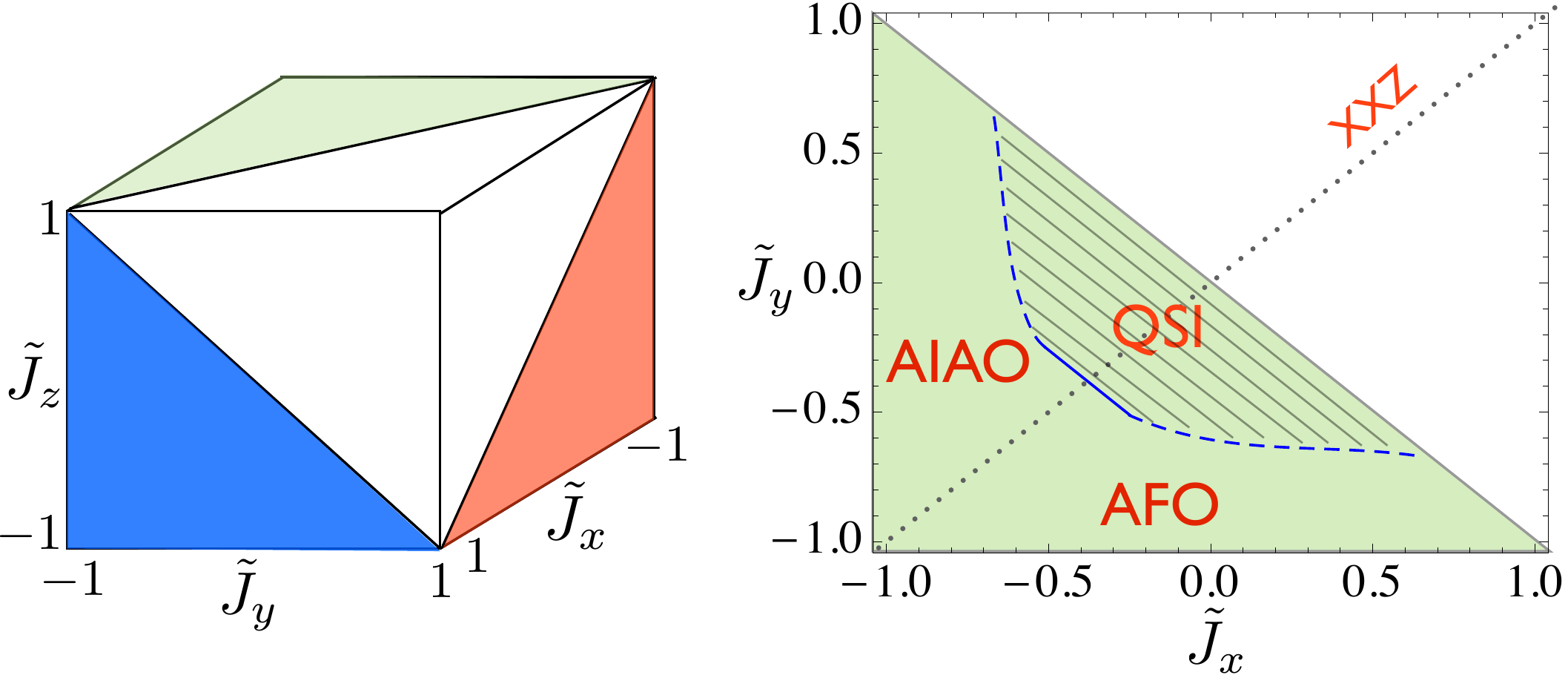} 
	\caption{(Color online). 
Left: Unit cube in $(\tilde{J}_x, \tilde{J}_y, \tilde{J}_z)$ parameter space of the XYZ model.  Shaded regions were analyzed via gMFT.  
Right: gMFT phase diagram on the $\tilde{J}_z =1$ surface of the cube, where dQSI, all-in-all-out (AIAO), and anti-ferro-octupolar (AFO) phases are found.  Within gMFT, the phase transition is 1st order 
(2nd order) at the dashed (solid) boundary. The dotted line is the XXZ line.  We did not apply gMFT for $\tilde{J}_x + \tilde{J}_y \geq 0$. There, the exchange is frustrated, and QSI is likely to be more stable than for $\tilde{J}_x + \tilde{J}_y < 0$ \cite{Lee2012}.  The phase diagram on the other surfaces of the cube can be obtained by relabeling parameters, with the nature of phases changing according to the anisotropic character of $\tilde{\tau}^{\mu}_{\br}$.  dQSI occurs on the $\tilde{J}_z = 1$ and $\tilde{J}_x = 1$ faces, while oQSI occurs on the $\tilde{J}_y = 1$ face.}
	\label{fig:fig3}
\end{figure}

So far, we have avoided discussing the case $\tilde{J}_{\perp} > 0$; here, less is known for the XXZ model, due to the presence of a sign problem in quantum Monte Carlo. In the $|\tilde{J}_\perp|  / \tilde{J}_z \ll 1$ limit, $\tilde{J}_\perp$ favors QSI with $\pi$ flux of the vector potential $A_{\bmr \bmr'}$ through each pyrochlore hexagon, unlike for $\tilde{J}_\perp < 0$, where zero flux is favored \cite{Note1}.  We have not considered the properties of the resulting $\pi$-flux versions of dQSI and oQSI, leaving this for future work. 
QSI is expected to persist over a larger range of $\tilde{J}_\perp > 0$, since in this case both $\tilde{J}_z$ and $\tilde{J}_\perp$ interactions are frustrated \cite{Lee2012}.

We now discuss the phase diagram of the XYZ model.  The simplest magnetically ordered phases appear ferromagnetic in local coordinates; for instance, if $\tilde{J}_z < 0$ and is dominant, $\langle \tilde{\tau}^z_{\br} \rangle  = m_d \neq 0$.  This is the ``all-in-all-out'' (AIAO)  state, where dipoles point along the local $z_i$ axes, toward (away from) pyrochlore tetrahedron centers lying in the diamond A (B) sublattices (or vice versa).   Since $\tau^z$ and $\tau^x$ transform identically under space group, the same AIAO state  arises when $\tilde{J}_x < 0$, $|\tilde{J}_x| \gg \tilde{J}_{y,z}$.  A distinct magnetically ordered phase, with $\langle \tilde{\tau}^y_{\br} \rangle = m_o \neq 0$, is obtained when $\tilde{J}_y < 0$, $|\tilde{J}_y| \gg \tilde{J}_{x,z}$.  This state has anti-ferro-octupolar order, and no on-site dipolar order.

To study the phase diagram away from the simple limits discussed above, we employ gauge mean
field theory (gMFT)\cite{Savary2012,Lee2012} to our model\cite{Note1}.  gMFT makes the ${\rm U}(1)$ gauge structure explicit via a slave particle construction, and is capable of describing both QSI and magnetic phases.  
For simplicity, we limited our analysis to the shaded regions shown (Fig.~\ref{fig:fig3}) on the faces of a cube in $(\tilde{J}_x, \tilde{J}_y, \tilde{J}_z)$ space.
We find only the two QSI and magnetically ordered phases discussed above.  In the same regions of parameter space we analyzed via gMFT, the XYZ model can be studied via quantum Monte Carlo without a sign problem \cite{Note1}.

We now comment on the prospects for applying the models discussed above to real materials.  Promising systems to realize the XYZ model are Nd$_2$B$_2$O$_7$ pyrochlores.  ${\rm B} = {\rm Zr}, {\rm Sn}$ compounds are insulators exhibiting  antiferromagnetic order at low temperature \cite{Blote1969,Matsuhira2002}.  While the ${\rm B} = {\rm Ir}$ compound is known to carry a DO doublet \cite{Watahiki2011}, the physics is  complicated by the presence of Ir conduction electrons\cite{Chen2012}.  Synthesis of other Nd pyrochlores has been reported  \cite{Subramanian1983}.  The validity of the XYZ model description could be ascertained and the exchange couplings measured directly via neutron scattering in applied magnetic field, as was done in the dipolar case for Yb$_2$Ti$_2$O$_7$ \cite{Ross2011}.  DO doublets are likely in Dy pyrochlores  \cite{Bertin2012}, but the large moment of Dy$^{3+}$ means dipolar interactions must be included.  DO doublets may also occur in B-site rare earth spinels, and there is evidence for this in CdEr$_2$Se$_4$ \cite{Lago2010}.   More broadly, strongly localized $d$-electron Mott insulators with $S = 3/2$ and $D_{3d}$ site symmetry comprise another class of systems where DO doublets may be the low-energy degrees of freedom.

$5d$ systems are a likely setting for itinerant (or weakly localized) DO doublets.  
Cd$_2$Os$_2$O$_7$, believed to exhibit AIAO order below a finite-temperature metal-insulator transition\cite{Koda2007,Yamaura2012}, has Os$^{3+}$ in  $5d^3$ configuration.  Microscopic calculations indicate a DO doublet ground state, but show a very small splitting between ground and first excited doublets \cite{Bogdanov2013}, likely due to Hund's coupling.  Moreover, electronic structure calculations do not show a clear separation between DO doublet and other energy bands \cite{Singh2002,Hiroshi2012}.  Thus $5d^1$ systems, perhaps on other lattices, may be more promising for the realization of itinerant DO doublets.  

In summary, we have pointed out that Kramers doublets with dipolar-octupolar character can arise on the sites of the pyrochlore lattice in both $d$- and $f$-electron systems.  We studied effective models of DO doublets in itinerant and localized limits, finding topological insulation in the former case, and two distinct quantum spin ice phases in the latter.

\emph{Acknowledgements.} -- We thank Leon Balents, Michel Gingras, Sungbin Lee and Lucile Savary 
for helpful conversations and correspondence. This work was supported by the U.S. Department of Energy, Office of Science, Basic Energy Sciences, under Award number DE-FG02-10ER46686.

\bibliographystyle{my}
\bibliography{XYZ_pyrochlore.bib}

\newpage

\appendix

\begin{center}
{\bf SUPPLEMENTARY MATERIAL}
\end{center}

\section{On-site interaction for $d^3$ electron configuration}
\label{sec:ssec1}

As discussed in the main text, for $d^3$ electron configuration, the on-site interaction plays an important role and must be included.
For a fixed lattice site, the interaction $H_{\text{int}}$ projected into the $t_{2g}$ manifold is of  Kanamori form,
\begin{eqnarray}
	H_{ \text{int} }&=&\frac{U}{2} \sum_m \sum_{ \sigma \neq \sigma'}
	d_{ m \sigma }^{\Da}d_{ m \sigma'}^{\Da}
         d^{\phantom\dagger}_{ m \sigma'}d^{\phantom\dagger}_{ m \sigma} 
         \nonumber \\
	&+&\frac{U'}{2}\sum_{m\neq m'}\sum_{\sigma,\sigma'}
	d_{ m \sigma}^{\Da}d_{ m' \sigma'}^{\Da}
         d^{\phantom\dagger}_{ m' \sigma'}d^{\phantom\dagger}_{ m \sigma}
	\nonumber\\
	&+&\frac{J}{2}\sum_{m\neq m'}\sum_{\sigma,\sigma'}
	d_{ m \sigma}^{\Da}d_{ m' \sigma'}^{\Da}
         d^{\phantom\dagger}_{ m \sigma'}d^{\phantom\dagger}_{ m' \sigma}
         \nonumber\\
	&+&\frac{J'}{2}\sum_{m\neq m'}
	\sum_{\sigma,\sigma'}d_{ m \sigma}^{\Da}d_{ m \sigma'}^{\Da}
         d^{\phantom\dagger}_{ m'\sigma'}d^{\phantom\dagger}_{ m'\sigma} \text{.}
\end{eqnarray}
Here, $d^{\dagger}_{m \sigma}$ creates an electron in the $t_{2g}$ orbital labeled by $m = 1,2,3$, with spin $\sigma = \uparrow, \downarrow$, and
$U,U',J,J'$ are the Kanamori parameters. 
For simplicity, we take the atomic limit by setting $U=U'+J+J'$ and $J=J'\equiv J_H$, where $J_H$ is the Hund's coupling.

\begin{figure}[t]
\includegraphics[width=7.5cm]{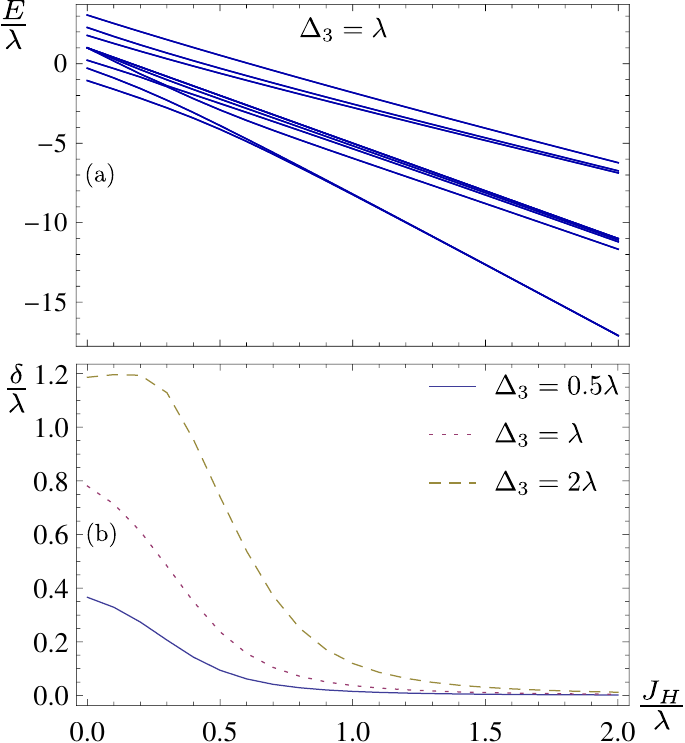}
\caption{(a) The energy spectrum of the single-site Hamiltonian for $d^3$ electron configuration, as a function of 
$J_H / \lambda$ at ${\Delta_3}=\lambda$.    (b)  Plot of the splitting $\delta$ between the first excited and ground doublets as a function of $J_H / \lambda$ for  three different values of $\Delta_3 / \lambda$.}
\label{fig:sfig1}
\end{figure}

We have assessed the effect of on-site interaction by direct diagonalization of the on-site Hamiltonian [Eq.~(1) in the main text], including spin-orbit coupling $\lambda$, trigonal crystal field splitting $\Delta_3$, as well as the interaction parameter $J_H$.  For a fixed number of electrons, the Hubbard interaction $U$ has no effect and can be neglected.  In Fig.~\ref{fig:sfig1}a, the energy spectrum is plotted as a function of $J_H / \lambda$ for $\Delta_3 = \lambda$.  The ground state is a DO doublet, and the first excitation is a dipolar doublet; we denote the splitting between these levels by $\delta$.  Independent of $J_H$, we find that the DO doublet remains the ground state ($\delta > 0$), but $\delta /\lambda$ becomes small for $J_H \gtrsim \lambda$, as large $J_H$ favors a $S = 3/2$ ground state.  The splitting $\delta$ is plotted in Fig.~\ref{fig:sfig1}b for different values of $\Delta_3 / \lambda$; it is apparent that larger trigonal splitting leads to larger separation between the two lowest doublets.

\section{Lattice geometry}
\label{sec:ssec2}

The pyrochlore lattice is a FCC lattice with four-site basis.  Setting the FCC lattice constant to unity, we choose the FCC primitive vectors to be
\begin{eqnarray}
\ba_1 &=& \frac{1}{2} (0, 1, 1) \\
\ba_2 &=& \frac{1}{2} (1, 0, 1) \nonumber \\
\ba_3 &=& \frac{1}{2} (1, 1, 0) \nonumber \text{.}
\end{eqnarray}
The basis vectors are taken to be
\begin{equation}
\boldsymbol{b}_i = - \frac{\sqrt{3}}{8} \hat{z}_i\text{,}
\end{equation}
where $\hat{z}_i$ is defined below, and $i = 1,\dots,4$ is the sublattice index.  The pyrochlore lattice can be viewed as composed of corner-sharing tetrahedra whose centers form a diamond lattice.  The A-sublattice diamond sites are $\{ \bR \}$, and the B-sublattice sites are $\{ \bR + \frac{1}{4}(1,1,1) \}$, where $\bR$ is an arbitrary FCC Bravais lattice vector.  The basis vectors themselves form the A-sublattice tetrahedron centered at the origin.  In the following we will use serif symbol $\boldsymbol{\mathsf{r}}, \boldsymbol{\mathsf{r}}'$ to label the sites on the dual diamond lattice and $\boldsymbol{r}, \boldsymbol{r}'$ to label the sites on the pyrochlore lattice.  Pyrochlore sites are also labeled by the pair $(\bR, i)$, where $\br = \bR + \boldsymbol{b}_i$.

It is convenient to introduce local
coordinate systems for each sublattice. These are given by unit vectors $(\hat{x}_i, \hat{y}_i, \hat{z}_i)$
defined as follows,
\begin{eqnarray}
	\hat{z}_1&=&\frac{1}{\sqrt{3}}(1,1,1)\quad\quad\quad
         \hat{y}_1=\frac{1}{\sqrt{2}}(0,1,-1), \nonumber\\
	\hat{z}_2&=&\frac{1}{\sqrt{3}}(1,-1,-1)\quad\,\,\,
         \hat{y}_2=\frac{1}{\sqrt{2}}(-1,0,-1), \nonumber\\
	\hat{z}_3&=&\frac{1}{\sqrt{3}}(-1,1,-1)\quad\,\,\,
         \hat{y}_3=\frac{1}{\sqrt{2}}(-1,-1,0), \nonumber\\
	\hat{z}_4&=&\frac{1}{\sqrt{3}}(-1,-1,1)\quad\,\,\,
         \hat{y}_4=\frac{1}{\sqrt{2}}(-1,1,0),
	 \label{localbasis}
\end{eqnarray}
and $\hat{x}_i \equiv \hat{y}_i\times\hat{z}_i$.  $\hat{z}_i$ the local  3-fold axis of the $D_{3d}$ site symmetry, and points toward the center of the A-sublattice tetrahedra.

\section{Symmetry analysis}

The $Fd\bar{3}m$ space group is generated by the following operations:  (1) Symmetries of the tetrahedron centered at $\sfr = 0$, forming the group $T_d$.
  (2) Inversion ${\cal I}$ about the site $\br = \boldsymbol{b}_1$.  (3) Primitive FCC translations  $T_{\ba_1}, T_{\ba_2}, T_{\ba_3}$.  We also consider time reversal symmetry ${\cal T}$.
  
The $T_d$ group preserving the $\sfr = 0$ tetrahedron is generated by $C_{3,1}$ and $M_{x \bar{y}}$.  Here, $C_{3,1}$ is a 3-fold rotation preserving the site $\br = \boldsymbol{b}_1$, and $M_{x \bar{y}}$ is a mirror reflection sending $x \leftrightarrow -y$.  Explicitly,
\begin{equation}
C_{3,1}: \br \to C_{3,1} \br \equiv \left( \begin{array}{ccc} 
0 & 0 & 1 \\
1 & 0 & 0 \\
0 & 1 & 0
\end{array} \right) \br \text{,}
\end{equation}
and
\begin{equation}
M_{x \bar{y} } : \br \to M_{x \bar{y}} \br  \equiv \left( \begin{array}{ccc} 
0 & -1 & 0 \\
-1 & 0 & 0 \\
0 & 0 & 1
\end{array} \right) \br \text{.}
\end{equation}

Below, we work out the effect of these symmetries on DO doublets, first for the simpler case of localized pseudospins, then for the case of itinerant electrons in DO doublets.

\subsection{Localized case}
\label{sec:localized}

For concreteness, we begin by considering $f$-electron magnetic moments on the sites of the pyrochlore lattice, with total angular momentum $J = 3/2, 9/2, 15/2$.  We suppose that crystal field splitting leads to a ground state DO doublet, with the \emph{same symmetry} as the doublet $J^{z_i} = \pm J$.  (Note that we do \emph{not} assume the ground state doublet is exactly given by $J^{z_i} = \pm J$, only that it transforms identically under symmetry.)
   Letting ${\cal P}$ project onto the $J^{z_i} = \pm J$ subspace, we define the pseudospin operators by
\begin{eqnarray}
\tau^z_{\bR i} &=& \frac{1}{2 J} {\cal P} J^{z_i}_{\bR i} {\cal P} \\
\tau^+_{\bR i} &=& \frac{1}{(2 J)!} {\cal P} (J^{+_i}_{\bR i})^{2J} {\cal P} \text{,}
\end{eqnarray}
where $\tau^-_{\bR i} = (\tau^+_{\bR i})^\dagger$, $\tau^{\pm}_{\bR i} = \tau^x_{\bR i} \pm i \tau^y_{\bR i} $, and $J^{\pm_i}_{\bR i} = J^{x_i}_{\bR i} \pm i J^{y_i}_{\bR i}$.  With these conventions, the pseudospin operator $\tau^{\mu}_{\bR i}$ ($\mu = x,y,z$) has eigenvalues $\pm 1/2$.  We can now proceed to determine the symmetry transformations of $\tau^{\mu}_{\bR i}$ in terms of the known transformations of $\bJ_{\bR i}$.

The above discussion applies directly to DO doublets obtained from $d$-electrons [case (1) in the main text], upon replacing $\bJ_{\bR i}$ with $\bj^{{\rm eff}}_{\bR i}$, and $J$ with $3/2$.  Both $\bJ_{\bR i}$ and $\bj^{{\rm eff}}_{\bR i}$ transform identically under symmetry, namely as  time-reversal odd pseudovectors.

The generators of the symmetry group act on $\bJ_{\br}$ as follows:
\begin{eqnarray}
T_{\ba_i} : \bJ_{\br} &\to& \bJ_{\br + \ba_i} \\
{\cal I} : \bJ_{\br} &\to& \bJ_{{\cal I} \br} \\
C_{3,1} : \bJ_{\br} &\to& C^T_{3,1} \bJ_{C_{3,1} \br} \\
M_{x \bar{y}} : \bJ_{\br} &\to& - M^T_{x \bar{y}} \bJ_{M_{x \bar{y}} \br} \\
{\cal T} : \bJ_{\br} &\to& - \bJ_{\br} \text{.}
\end{eqnarray}
For each symmetry operation $S$, we let $\hat{U}_S$ be the unitary operator representing it.
The above notation is short-hand for conjugation of $\bJ_{\br}$ by the appropriate unitary or anti-unitary operators representing each symmetry, \emph{e.g.} ${\cal I} : \bJ_{\br} \to \hat{U}^{\vphantom\dagger}_{\cal I} \bJ_{\br} \hat{U}^\dagger_{\cal I} = \bJ_{ {\cal I} \br}$.

From the above relations and the definition of $\tau^{\mu}_{\br}$, it is straightforward to show
\begin{eqnarray}
T_{\ba_i} : \tau^{\mu}_{\br} &\to& \tau^{\mu}_{\br + \ba_i} \\
{\cal I} : \tau^{\mu}_{\br} &\to& \tau^{\mu}_{{\cal I} \br} \\
C_{3,1} : \tau^{\mu}_{\br} &\to& \tau^{\mu}_{C_{3,1} \br} \\
M_{x \bar{y} } : \tau^{x,z}_{\br} &\to& - \tau^{x,z}_{M_{x \bar{y}} \br} \\
M_{x \bar{y} } : \tau^{y}_{\br} &\to&  \tau^{y}_{M_{x \bar{y}} \br} \\
{\cal T} : \tau^{\mu}_{\br} &\to& - \tau^{\mu}_{\br} \text{.} \label{eqn:T_on_pseudospin}
\end{eqnarray}
It is notable that $\tau^{\mu}_{\br}$ transforms trivially under $C_{3,1}$ and that $\tau^y_{\br}$ transforms trivially under \emph{all} space group operations.  This is a direct reflection of the octupolar character of $\tau^y$ (see Sec.~\ref{sec:donature}).

The space group transformation properties can be simply stated without choosing specific generators.  Consider the diamond lattice formed by the tetrahedron centers $\sfr$.  Every space group operation \emph{either} preserves the diamond A-sublattice (and hence also the B-sublattice), \emph{or} it exchanges A- and B-sublattices.  We refer to the former operations as A/B-preserving, and the latter as A/B-exchanging.  For improper A/B-preserving operations (\emph{e.g.} mirror planes), $\tau^{x,z}_{\br}$ is odd.  (More precisely, if $S$ is such an operation, then $S : \tau^{x , z}_{\br} \to - \tau^{x , z}_{S \br}$.)  For proper A/B-preserving operations, $\tau^{x,z}_{\br}$ is even.  This is reversed for A/B-exchanging operations, with $\tau^{x,z}$ even under improper operations and odd under proper operations.  Finally, $\tau^y_{\br}$ is even under \emph{all} space group operations.

\subsection{Itinerant case}
\label{sec:itinerant}

Here, we work out the effect of space group and time reversal symmetry on electron operators, as required to construct models of itinerant electrons in DO doublets.  Rather than pursuing a direct microscopic analysis, we adopt an indirect approach.  The idea is to first write down, for each generator, the most general transformation of the electron operator consistent with the pseudospin transformations derived above.  Each such transformation involves unknown phase factors, and in general the resulting transformations do not satisfy the relations defining the symmetry group.  
We show that, up to gauge transformations, the phase factors are \emph{completely} determined by requiring the group relations to hold.  

We let $c^\dagger_{\bR i \alpha}$, where $\alpha = \pm$, create an electron at site $(\bR, i)$ in pseudospin state $\tau^z_{\bR i} = 1/2$ for $\alpha = +$, and $\tau^z_{\bR i} = -1/2$ for $\alpha = -$.  It is convenient sometimes to suppress the pseudospin index and write $c^\dagger_{\bR i}$, which we can think of as a two-component row vector of operators.  Sometimes we suppress both spin and basis indices, writing $c^\dagger_{\bR}$, an 8-component row vector of operators.  The pseudospin operator is $\tau^{\mu}_{\bR i} = \frac{1}{2} c^\dagger_{\bR i} \sigma^{\mu} c^{\vphantom\dagger}_{\bR i}$.

Since translations $T_{\ba_i}$ commute, a gauge can be chosen in which
\begin{equation}
T_{\ba_i} : c^\dagger_{\bR} \to \hat{U}^{\vphantom\dagger}_{T_{\ba_i}} c^\dagger_{\bR}\hat{U}^{\dagger}_{T_{\ba_i}}  = c^\dagger_{\bR + \ba_i} \text{.}
\end{equation}
The residual gauge freedom preserving this form of $T_{\ba_i}$ is $c^\dagger_{\bR i} \to \alpha^g_i c^\dagger_{\bR i}$, where $\alpha^g_i \in {\rm U}(1)$.

The most general action of time reversal consistent with Eq.~(\ref{eqn:T_on_pseudospin}) is ${\cal T} : c^\dagger_{\bR} \to  
{\cal T} c^\dagger_{\bR} {\cal T}^{-1} = c^\dagger_{\bR} U^{\cal T}_{\bR}$, where
\begin{equation}
U^{\cal T}_{\bR} = \left( \begin{array}{cccc}
\alpha^{\cal T}_{\bR 1} (i \sigma^y) & 0 & 0 & 0 \\
0 & \alpha^{\cal T}_{\bR 2} (i \sigma^y) & 0 & 0 \\
0 & 0 & \alpha^{\cal T}_{\bR 3} (i \sigma^y) & 0 \\
0 & 0 & 0 & \alpha^{\cal T}_{\bR 4} (i \sigma^y) 
\end{array} \right) \text{,}
\end{equation}
where $\alpha^{\cal T}_{\bR i} \in {\rm U}(1)$.  Using the fact that $\hat{U}_{T_{\ba_i}} {\cal T} = {\cal T} \hat{U}_{T_{\ba_i}}$, it is easy to show $\alpha^{\cal T}_{\bR i} \equiv \alpha^{\cal T}_i$.  Moreover, we can make a gauge transformation to set $\alpha^{\cal T}_i = 1$, and thus ${\cal T} : c^\dagger_{\bR i} \to c^\dagger_{\bR i} (i \sigma^y)$.  The residual gauge freedom preserving the form of both translations and time reversal is still  $c^\dagger_{\bR i} \to \alpha^g_i c^\dagger_{\bR i}$, but now each $\alpha^g_i \in \{\pm 1 \}$.

The most general form of $C_{3,1}$ rotation consistent with the pseudospin transformations is 
\begin{equation}
C_{3,1} : c^\dagger_{\bR} \to \hat{U}^{\vphantom\dagger}_{C_{3,1}} c^\dagger_{\bR}  \hat{U}^{\dagger}_{C_{3,1}} = c^\dagger_{C_{3,1} \bR} U^{C_{3,1}}_{\bR} \text{,}
\end{equation}
 where the $8 \times 8$ matrix $U^{C_{3,1}}_{\bR}$ is given in $2 \times 2$ block form by
\begin{equation}
U^{C_{3,1}}_{\bR} = \left( \begin{array}{cccc}
\alpha^C_{\bR 0} & 0 & 0 & 0 \\
0 & 0 & 0 & \alpha^C_{\bR 1} \\
0 & \alpha^C_{\bR 2} & 0 & 0 \\
0 & 0 & \alpha^C_{\bR 3} & 0
\end{array}\right) \text{,}
\end{equation}
for $\alpha^C_{\bR i} \in {\rm U}(1)$.  This is simplified by noting that $\hat{U}_{C_{3,1}} \hat{U}_{T_{\ba_i}} = \hat{U}_{T_{C_{3,1} \ba_i}} \hat{U}_{C_{3,1}}$ implies $\alpha^C_{\bR i} \equiv \alpha^C_i$, and ${\cal T} \hat{U}_{C_{3,1}} 
=\hat{U}_{C_{3,1}}  {\cal T}$ gives $\alpha^C_i \in \{ \pm 1 \}$.

To proceed further, we employ the relation $[\hat{U}_{C_{3,1}}]^3 = -1$, where the minus sign reflects the $S=1/2$ nature of electrons.  This implies that $\alpha^C_1 = -1$, and $\alpha^C_2 \alpha^C_3 \alpha^C_4 = -1$.  It is then possible set all $\alpha^C_i = -1$, by making a gauge transformation of the form $\alpha^g_1 = 1$, and $\alpha^g_{i} \in \{ \pm 1 \}$ (for $i=2,3,4$).  The resulting form of $C_{3,1}$ rotation is still preserved by gauge transformations with $\alpha^g_1 \in \{ \pm 1\}$ and $\alpha^g_i = 1$ (for $i=2,3,4$).

Next we consider the mirror reflection $M_{x \bar{y}}$, which acts on electron operators by
\begin{equation}
\mxyb : c^\dagger_{\bR} \to \hat{U}^{\vphantom\dagger}_{\mxyb} c^\dagger_{\bR}  \hat{U}^{\dagger}_{\mxyb} = c^\dagger_{\mxyb \bR} U^{\mxyb}_{\bR} \text{,}
\end{equation}
where
\begin{equation}
U^{\mxyb} = \left( \begin{array}{cccc}
0 & 0 & 0 & \alpha^M_{\bR 1} (i \sigma^y) \\
0 &  \alpha^M_{\bR 2} (i \sigma^y) & 0 & 0 \\
0 & 0 &  \alpha^M_{\bR 3} (i \sigma^y) & 0 \\
 \alpha^M_{\bR 4} (i \sigma^y) & 0 & 0 & 0 
 \end{array}\right) \text{,}
 \end{equation}
 where $\alpha^M_{\bR i} \in {\rm U}(1)$.  The relation $\hat{U}_{\mxyb} \hat{U}_{T_{\ba_i}} = \hat{U}_{T_{\mxyb \ba_i}} \hat{U}_{\mxyb}$ implies $\alpha^M_{\bR i} \equiv \alpha^M_i$, and  ${\cal T} \hat{U}_{\mxyb} 
=\hat{U}_{\mxyb}  {\cal T}$ gives $\alpha^M_i \in \{ \pm 1 \}$.  

Viewing $\mxyb$ as the composition of a $C_2$ rotation and an inversion, we require the relation $[ \hat{U}_{\mxyb}]^2 = -1$, which implies $\alpha^M_1 = \alpha^M_4$.  The relation $[ \hat{U}_{\mxyb} \hat{U}_{C_{3,1}} ]^4 = -1$ then implies $\alpha^M_2 = - \alpha^M_3$.  To fix the remaining free parameters, in addition to gauge freedom, we have the freedom to redefine $\hat{U}_{\mxyb} \to - \hat{U}_{\mxyb}$, which allows us to set $\alpha^M_2 = -1$, $\alpha^M_3 = 1$.  We can then make a gauge transformation of the form $\alpha^g_1 \in \{ \pm 1 \}$, $\alpha^g_i = 1$ (for $i=2,3,4$), to set $\alpha^M_1 = \alpha^M_4 = 1$, thus completely fixing the form of $\hat{U}_{\mxyb}$.

The only remaining generator is inversion, which acts on electron operators by
\begin{equation}
{\cal I} : c^\dagger_{\bR i} \to \hat{U}^{\vphantom\dagger}_{\cal I} c^\dagger_{\bR i} \hat{U}^{\dagger}_{\cal I} 
= c^\dagger_{[-\bR + 2(\bb_1 - \bb_i)],i} \alpha^{\cal I}_{\bR i} \text{,}
\end{equation}
for $\alpha^{\cal I}_{\bR i} \in {\rm U}(1)$.  The relation $\hat{U}_{T_{\ba_i}} \hat{U}_{\cal I} = \hat{U}_{\cal I} T_{- \ba_i}$ implies
$\alpha^{\cal I}_{\bR i} \equiv \alpha^{\cal I}_i$, and ${\cal T} \hat{U}_{\cal I} = \hat{U}_{\cal I} {\cal T}$ gives $\alpha^{\cal I} \in \{ \pm 1 \}$.

To proceed further, considering the action of the relation $\hat{U}_{\cal I} \hat{U}_{C_{3,1}} = \hat{U}_{C_{3,1}} \hat{U}_{\cal I}$ on $c^\dagger_{\bR i}$, for $\bR = 0$, gives $\alpha^{\cal I}_2 = \alpha^{\cal I}_3 = \alpha^{\cal I}_4$.  Similarly, acting on $c^\dagger_{0,0}$ with $\hat{U}_{T_{\ba_3}} \hat{U}_{\cal I} \hat{U}_{\mxyb} = \hat{U}_{\mxyb} \hat{U}_{\cal I}$  gives $\alpha^{\cal I}_1 = \alpha^{\cal I}_4$, and thus all the $\alpha^{\cal I}_i$ are equal. We can then set $\alpha^{\cal I}_i = 1$ by exploiting the freedom to send $\hat{U}_{\cal I} \to - \hat{U}_{\cal I}$.

We have thus completely fixed the action of symmetry on electron operators.  To summarize, we have obtained the following results:
\begin{eqnarray}
T_{\ba_i} : c^\dagger_{\bR} &\to& c^\dagger_{\bR + \ba_i} \\
{\cal I} : c^\dagger_{\bR i} &\to& c^\dagger_{[-\bR + 2(\bb_1 - \bb_i)],i} \\
C_{3,1} : c^\dagger_{\bR} &\to& c^\dagger_{C_{3,1} \bR} U^{C_{3,1}} \\
\mxyb : c^\dagger_{\bR} &\to& c^\dagger_{\mxyb \bR} U^{\mxyb} \\
{\cal T} : c^\dagger_{\bR i} &\to& c^\dagger_{\bR i} ( i \sigma^y) \text{,}
\end{eqnarray}
where the $8 \times 8$ matrices $U^{C_{3,1}}$ and $U^{\mxyb}$ are given in $2 \times 2$ block form by
\begin{equation}
U^{C_{3,1}}_{\bR} = \left( \begin{array}{cccc}
-1 & 0 & 0 & 0 \\
0 & 0 & 0 & -1 \\
0 & -1 & 0 & 0 \\
0 & 0 & -1 & 0
\end{array}\right) \text{,}
\end{equation}
\begin{equation}
U^{\mxyb} = \left( \begin{array}{cccc}
0 & 0 & 0 & (i \sigma^y) \\
0 &  - (i \sigma^y) & 0 & 0 \\
0 & 0 &   (i \sigma^y) & 0 \\
 (i \sigma^y) & 0 & 0 & 0 
 \end{array}\right) \text{.}
 \end{equation}

We have also obtained the same results directly from a microscopic analysis\cite{MicroUnpub}, but prefer the indirect approach presented here both for its greater technical simplicity, and the additional insight it provides.

\section{Dipolar-octupolar nature of the doublets}
\label{sec:donature}

Here, we consider the transformation of $\tau^{\mu}$ under the $D_{3d}$ site symmetry.  We show that $\tau^{x,z}$ transform like $m_{z}$, the $z$-component of a magnetic dipole, while $\tau^y$ does not transform like any component of a magnetic dipole.  Instead, $\tau^y$ transforms like a component of the magnetic octupole tensor $T_{\mu \nu \lambda}$.  In this section, we will consider a fixed site $\br$.  We consider dipole $m_{\mu}$ and octupole $T_{\mu \nu \lambda}$ tensors in the local coordinates introduced in Sec.~\ref{sec:ssec2}, suppressing the basis index $i$ to simplify the notation.

The $D_{3d}$ site symmetry is generated by 3-fold rotation $C_3$, inversion ${\cal I}$, and a mirror plane $M$.  (There are three different mirror planes; we arbitrarily choose one of these to be a generator.)  The dipole $m_{\mu}$ transforms as a pseudovector under these operations, as does each index of $T_{\mu \nu \lambda}$.  It follows from Sec.~\ref{sec:localized} that
\begin{eqnarray}
C_3 : \tau^{\mu} &\to& \tau^{\mu} \\
M : \tau^{x,z} &\to& - \tau^{x,z} \\
M : \tau^y &\to& \tau^y \\
{\cal I} : \tau^{\mu} &\to& \tau^{\mu} \text{.}
\end{eqnarray}
It is clear that $\tau^{\mu}$ does not transform as a pseudovector.  We observe that $\tau^z$ and $\tau^x$ transform identically to one another, and also to $m_z$; therefore the $\mu = x,z$ components of $\tau^{\mu}$ are dipolar.  On the other hand, $\tau^y$ does not transform as any component of $m_{\mu}$.  Note that inversion does not play an important role, acting trivially on  $\tau^{\mu}$, $m_{\mu}$ and $T_{\mu \nu \lambda}$.

Instead, $\tau^y$ transforms identically to a component of $T_{\mu \nu \lambda}$.  To identify the appropriate component, we change coordinates from $\mu = x,y,z$ to $\alpha = +, -, z$ by writing
\begin{eqnarray}
m_{\alpha} &=& R_{\alpha \mu} m_{\mu} \\
T_{\alpha \beta \gamma} &=& R_{\alpha \mu} R_{\beta \nu} R_{\gamma \lambda} T_{\mu \nu \lambda} \text{,}
\end{eqnarray}
where
\begin{equation}
R = \left( \begin{array}{ccc}
1 & i & 0 \\
1 & -i & 0 \\
0 & 0 & 1 
\end{array}\right) \text{.}
\end{equation}
We have transformation laws
\begin{eqnarray}
C_3 : m_{\pm} &\to& e^{\pm 2 \pi i /3} m_{\pm} \\
C_3 : m_z &\to& m_z \\
M : m_{\pm} &\to&  e^{\pm i \phi_M} m_{\mp} \\
M : m_z &\to&  - m_z \text{,}
\end{eqnarray}
with the same transformations holding for each index of $T_{\alpha \beta \gamma}$.  Here, $\phi_M = \pi, \pi/3, -\pi/3$, depending on which of the three $D_{3d}$ mirror planes is chosen for $M$.  For our purposes, the phase factor $\phi_M$ is unimportant, as it drops out in the transformation of $T_{+++}$ and $T_{---}$; that is
\begin{eqnarray}
M : T_{+++} &\to& - T_{---} \\
M : T_{---} &\to& - T_{+++} \text{.}
\end{eqnarray}

Using these transformations, we can identify
\begin{equation}
\tau^y  \sim i ( T_{+++} - T_{---} ) \text{.}
\end{equation}
As desired, the right-hand side is real and time-reversal odd, since time reversal ${\cal T} : m_{\pm} \to - m_{\mp}$.

\section{Tight-binding model}
\label{sec:ssec3}

Here we describe the symmetry allowed tight-binding model for itinerant 
electrons in DO doublets on the pyrochlore lattice, and provide more information on 
the analysis of the corresponding electron band structures.

\begin{figure}[t]
\includegraphics[width=7.5cm]{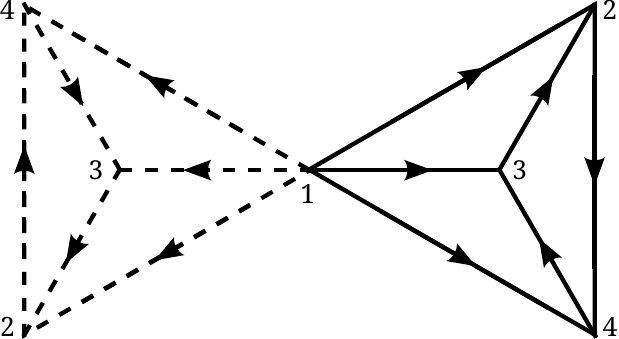}
\caption{Orientations of nearest-neighbor bonds for which the nearest-neighbor hopping takes an identical form on every bond.  The sites are numbered by basis index $i = 1,\dots,4$.  The center of the tetrahedron on the right (solid line bonds) lies in the diamond A-sublattice, while that of the left-hand tetrahedron (dashed-line bonds) lies in the diamond B-sublattice.}
\label{fig:bond-orientations}
\end{figure}

Requiring $Fd\bar{3}m$ space group and time reversal symmetry, the symmetry transformations given in Sec.~\ref{sec:itinerant} can be used to determine the most general tight-binding model allowed by symmetry.
The electron hopping has the general form given in Eq.~(2) of the main text
\begin{equation}
	H_{TB} = \sum_{ ( \bsr , \bsr'  )}\left[
	c_{\bsr}^{\Da}  T_{\bsr \bsr'}
        c^{\phantom\dagger}_{\bsr'}+h.c. \right]\text{,} 
	\label{B1}
\end{equation}
where the sum is over all bonds, with some arbitrary but fixed choice of orientation for each bond.  Time reversal symmetry implies $T_{\bsr \bsr'} = t^0_{\bsr \bsr'} + i t^{\mu}_{\bsr \bsr'} \sigma^\mu$.  For nearest-neighbor bonds $\langle \bsr \bsr' \rangle$, space group symmetry implies
\begin{equation}
T_{\langle \bsr \bsr' \rangle}  = i t^1_{nn} \sigma^x  + i t^3_{nn} \sigma^z \text{,}
\end{equation}
with the orientation shown in Fig.~\ref{fig:bond-orientations}.  The nearest-neighbor hopping Hamiltonian thus has an \emph{identical} form on every bond.
A pseudospin rotation about the local $y_i$ axes can eliminate one 
hopping parameter, resulting in 
\begin{equation}
	H_{nn} = \sum_{\langle \bsr \bsr' \rangle} \tilde{c}^{\Da}_{\bsr}
         (i\tilde{t}^3_{nn}\sigma^z) \tilde{c}_{\bsr'}^{\phantom\dagger}+h.c.\text{,} \label{eqn:simplenn}
\end{equation}
where $\tilde{t}^3_{nn} = \sqrt{ (t_{nn}^1)^2 + (t_{nn}^3)^2  }$
and $\tilde{c}^{\phantom\dagger}_{\bsr}, \tilde{c}^{\dagger}_{\bsr}$ are
electron operators in the rotated basis.

\begin{figure}[htp]
	\centering
	\includegraphics[width=8.5cm]{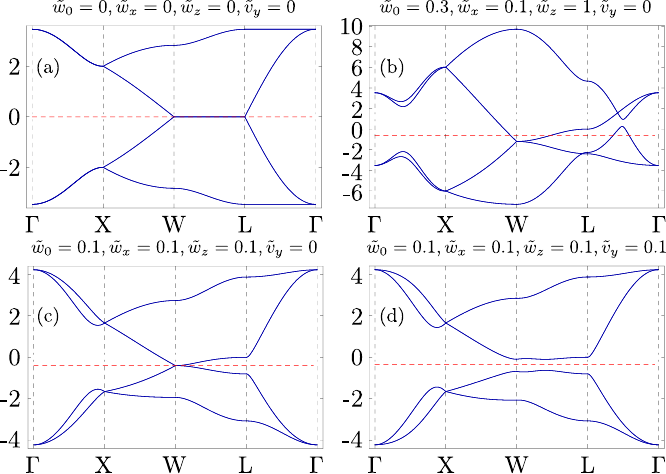}
	\caption{Band structure along high symmetry lines with energy in
	units of $\tilde{t}^3_{nn}$. (a) Only
	nearest-neighbor hopping is considered. (b) The metallic phase when
	both nearest-neighbor and second-neighbor hoppings are
	present. (c) The semimetal phase with W-point band touchings when nearest-neighbor
         and second-neighbor hoppings are present. (d)
         The strong topological insulator phase after the fourth-neighbor hopping is included on 
         top of the nearest-neighbor and second-neighbor hopping. 
         The topological invariant of the strong topological insulator is 
         $(\nu;\nu_1,\nu_2,\nu_3)=(1;0,0,0)$.}
	\label{fig:sfig2}
\end{figure}

The nearest-neighbor model at half-filling has a non-generic and highly unstable Fermi surface
specified by $\cos \frac{k_x}{2}+\cos \frac{k_y}{2}+\cos \frac{k_z}{2}=0$, which coincides with a surface of intersection between two bands.  The corresponding band structure is plotted in Fig.~\ref{fig:sfig2}(a). This highly fine-tuned 
Fermi surface is unstable to further-neighbor hopping, and it is thus crucial to include at least second-neighbor hopping in the tight-binding model.

Letting $(\bsr \bsr')_2$ label second-neighbor bonds, we specify the second-neighbor hopping by giving $T_{\bsr \bsr'}$ on a reference bond $(\bsr_0 \bsr'_0)_2$, where $\bsr_0 = \boldsymbol{b}_2$,  $\bsr_0' = -\frac{1}{4}(1,1,1)-\boldsymbol{b}_4$.  We have
\begin{equation}
T_{(\bsr_0 \bsr'_0)_2} = w_0 + i w_x \sigma^x + i w_z \sigma^z \text{,}
\end{equation}
where $w_y$ is forbidden by the $C_2$ rotation symmetry taking the bond into itself.  The reference bond can be mapped into any second-neighbor bond by an appropriate space group operation, so all $T_{(\bsr \bsr')_2}$ can be obtained from $T_{(\bsr_0 \bsr'_0)_2}$.  Unlike for nearest-neighbor hopping, the form of $T_{(\bsr \bsr')_2}$ varies from bond to bond.  The global pseudospin rotation resulting in Eq.~(\ref{eqn:simplenn}) for nearest-neighbor hopping affects the second-neighbor hopping merely by transforming the parameters $(w_0, w_x, w_z) \to (\tilde{w}_0, \tilde{w}_x, \tilde{w}_z)$.
 
Including second-neighbor hopping, we
find that the ground state 
is either a metal [Fig.~\ref{fig:sfig2}(b)] or semimetal with isolated  four-fold band touchings at the W-points.
[Fig.~\ref{fig:sfig2}(c)]. The phase diagram is discussed in the main text.  The putative semi-metal phase is an incipient  topological band insulator.  Because there is a gap at all time reversal invariant momenta, the $Z_2$ invariant can be computed using the Fu-Kane formula \cite{FuKane2007}, and is found to correspond to a strong topological insulator.
 This implies that \emph{any} time reversal preserving perturbation that opens a full gap leads to a strong topological insulator.

In fact, the W point band touching in the semimetal phase is unstable, and its presence is an artifact of restriction to only first- and second-neighbor hopping.  Upon including fourth-neighbor hopping, a gap opens at the W point, resulting in a strong topological insulator (third-neighbor hopping does not open a gap).
To establish this, among the 6 distinct W points, we focus on $\bk_W = (2\pi, \pi, 0)$.  Letting $H(\bk)$ be the $8 \times 8$ Bloch Hamiltonian including first- and second-neighbor hopping, we observe that $H(\bk_W)$ is block-diagonalized by the unitary transformation $\tilde{H}(\bk) = U^\dagger_W H(\bk) U_W$, where
\begin{equation}
U_W = \frac{1}{\sqrt{2}} \left( \begin{array}{cccc} 1 & 0 & 1 & 0 \\
0 & 1 & 0 & 1 \\
i & 0 & -i & 0 \\
0 & i & 0 & -i
\end{array}\right) \otimes {\mathbbm 1}_{2 \times 2}  \text{.}
\end{equation}
Here ${\mathbbm 1}_{2 \times 2}$ is the identity matrix acting in the DO doublet pseudospin space.  We find, in $4 \times 4$ block form
\begin{equation}
\tilde{H}(\bk_W) = \left( \begin{array}{cc}
\epsilon_F {\mathbbm 1}_{4 \times 4} & 0 \\
0 & K
\end{array}\right) \text{,}
\end{equation}
where the upper-left block acts in the manifold of the 4-fold touching, $\epsilon_F$ is the Fermi energy, and $K$ is a Hermitian $4 \times 4$ matrix with eigenvalues not equal to $\epsilon_F$.

To proceed, we consider the $4 \times 4$ effective Hamiltonian $H_{{\rm eff}}(\bq)$ that describes the splitting of the band touching for $\bk = \bk_W + \bq$, where $\bq$ is small compared to the Brillouin zone size.  In principle, this can be constructed by expanding the Bloch Hamiltonian $H(\bk_W + \bq)$ in powers of $\bq$, and treating the $\bq$-dependent terms via degenerate perturbation theory.  

For the present purposes, it is more useful to determine the most general form of $H_{{\rm eff}}(\bq)$ allowed by symmetry.  The group of the wavevector for $\bk_W$ is isomorphic to $C_{4v}$, and is generated by the four-fold rotation-reflection $S_{4y} = C_{3,1} \mxyb$ and the mirror reflection $M_x = {\cal I} C_{2x}$, where $C_{2x} \in T_d$ is a $\pi$-rotation about the $(100)$ axis.  In addition, the composition of inversion and time reversal ${\cal I} {\cal T}$ is an anti-unitary symmetry leaving all $\bk$-points invariant.
Using the results of Sec.~\ref{sec:itinerant}, and the transformation $U_W$, we determined the action of these symmetries in the 4-dimensional manifold of the band touching.  To quote the results, we introduce the operators $T^{\mu} = \sigma^{\mu} \otimes {\mathbbm 1}_{2 \times 2}$ and $\Sigma^{\mu} = {\mathbbm 1}_{2 \times 2} \otimes \sigma^{\mu}$; any $4 \times 4$ Hermitian matrix can be written as a real linear combination of ${\mathbbm 1}_{4 \times 4}$, $T^{\mu}$, $\Sigma^{\mu}$, and $T^{\mu} \Sigma^{\nu}$.  The symmetries act as follows:
\begin{eqnarray}
S_{4y} : \Sigma^{x,z} &\to& - \Sigma^{x,z} \\
S_{4y} : \Sigma^y &\to& \Sigma^y \\
S_{4y} : T^x &\to& - T^y \\
S_{4y} : T^y &\to& - T^x \\
S_{4y} : T^z &\to& - T^z \\
S_{4y} : (q_x, q_y, q_z) &\to& (q_z, -q_y, -q_x) \text{,}
\end{eqnarray}
and
\begin{eqnarray}
M_x : \Sigma^{\mu} &\to& \Sigma^{\mu} \\
M_x : T^x &\to& - T^y \\
M_x : T^y &\to& - T^x \\
M_x : T^z &\to& - T^z \\
M_x : (q_x, q_y, q_z) &\to& (-q_x, q_y, q_z) \text{,}
\end{eqnarray}
and finally
\begin{eqnarray}
{\cal I} {\cal T} : \Sigma^{\mu} &\to& - \Sigma^{\mu} \\
{\cal I} {\cal T} : T^x &\to& T^y \\
{\cal I} {\cal T}  : T^y &\to& T^x \\
{\cal I} {\cal T}  : T^z &\to& T^z \\
{\cal I} {\cal T} : \bq &\to& \bq \text{.}
\end{eqnarray}

The most general Hamiltonian respecting ${\cal I} {\cal T}$ is
\begin{equation}
H_{{\rm eff}}(\bq) = \epsilon_0(\bq) {\mathbbm 1}_{4 \times 4} +  f_a(\bq) \gamma_a \text{,}
\end{equation}
where $a = 1,\dots,5$, the $f_a(\bq)$ are arbitrary functions of $\bq$, and
\begin{eqnarray}
\gamma_1 &=& \frac{1}{\sqrt{2}} \Sigma^x (T^x - T^y) \\
\gamma_2 &=& \frac{1}{\sqrt{2}} \Sigma^y (T^x - T^y) \\
\gamma_3 &=& \frac{1}{\sqrt{2}} \Sigma^z (T^x - T^y) \\
\gamma_4 &=& \frac{1}{\sqrt{2}} ( T^x + T^y) \\
\gamma_5 &=& T^z \text{.}
\end{eqnarray}
The $\gamma_a$ matrices satisfy the $\gamma$-matrix algebra  $\{ \gamma_a, \gamma_b \} = 2 \delta_{a b}$, and it follows that the energy spectrum of $H_{{\rm eff}}(\bq)$ is
\begin{equation}
E_{\pm}(\bq) = \epsilon_0(\bq)  \pm \sqrt{ \sum_{a = 1}^5 [ f_a(\bq)]^2 } \text{,}
\end{equation}
where each (non-zero) energy level is two-fold degenerate.  A band touching occurs at $\bq$ only when $f_a(\bq) = 0$ for all $a$.  In the putative W-point semi-metal, there is an isolated touching at $\bq = 0$, and $\epsilon_0(0) = \epsilon_F$.

The remaining symmetries $S_{4y}$ and $M_x$ constrain the form of the $f_a(\bq)$.  Keeping terms up through second order in $\bq$, we find:
\begin{eqnarray}
f_1(\bq) &=& c_{1y} q_y + c_{1 xx} (q_x^2 - q_z^2) \\
f_2(\bq) &=& m + c_{2 yy} q^2_y + c_{2 xx} (q^2_x + q^2_z) \\
f_3(\bq) &=& c_{3y} q_y + c_{3 xx} (q_x^2 - q_z^2) \\
f_4(\bq) &=& c_{4 x z} q_x q_z \\
f_5(\bq) &=& c_{5 x z} q_x q_z \\
\epsilon_0(\bq) &=& \epsilon_F + e_{0 yy} q^2_y + e_{0 xx} (q^2_x + q^2_z) \text{.}
\end{eqnarray}

There is clearly a 4-fold touching at $\bq = 0$ only if $m = 0$, so evidently it happens that $m$ vanishes if we only include first- and second-neighbor hopping.  This can be understood by recalling that first- and second-neighbor hopping do not involve the $\sigma^y$ Pauli matrix in the DO doublet space, and thus the $8 \times 8$ Bloch Hamiltonian can be written in the form
\begin{equation}
H(\bk) = M_0(\bk) \otimes {\mathbbm 1}_{2 \times 2} + M_1(\bk) \otimes \sigma^x + M_3(\bk) \otimes \sigma^z \text{,}
\end{equation}
where the $M_{0,1,3}(\bk)$ are $4 \times 4$ Hermitian matrices. It follows from the form of $U_W$ that also
\begin{equation}
\tilde{H}(\bk) = \tilde{M}_0(\bk) \otimes {\mathbbm 1}_{2 \times 2} + \tilde{M}_1(\bk) \otimes \sigma^x + \tilde{M}_3(\bk) \otimes \sigma^z
\end{equation}
Now, $H_{{\rm eff}}(\bq = 0)$ is simply the upper-left $4 \times 4$ block of $\tilde{H}(\bk_W)$, and we thus see that $\Sigma^y$ and $\Sigma^y T^{\mu}$ terms cannot appear.  In particular, since $\gamma_2 = \Sigma^y (T^x - T^y)/\sqrt{2}$, this implies that $f_2(0) = 0$.  This result holds unless we consider hopping that involves $\sigma^y$ in the DO doublet space; it turns out that the shortest-range hopping for which this occurs is fourth-neighbor.

Now we consider the effect of a small $m \neq 0$ on the energy spectrum.  We are interested in the presence or absence of a gap.  We first note that, in the quadratic approximation for $f_a(\bq)$, and for generic values of the parameters $c_{1y}, c_{1 xx}$, etc., $\sum_{a \neq 2} [f_a(\bq)]^2 \neq 0$ for $\bq \neq 0$.  Then since $f_2(0) = m$, $\sum_a [f_a(\bq)]^2 \neq 0$ for all $\bq$, and a full gap is opened.  For generic parameters, this result is also expected to hold beyond the quadratic approximation, since each $f_a(\bq) = 0$ defines a surface in $\bq$-space, and the four surfaces for $a = 1,3,4,5$, apart from intersecting at $\bq = 0$, are not expected to have other intersections.  

The above discussion implies that a small fourth-neighbor hopping will open a gap at the W point, which we have directly verified.  A band structure illustrating this effect is shown in Fig.~\ref{fig:sfig2}(d).

\section{XYZ model and quantum spin ice}
\label{sec:ssec4}

Here, we mention some features of the XYZ Hamiltonian
\begin{equation}
H_{\text{XYZ}} = \sum_{\langle \br \br' \rangle}
	\tilde{J}_{x} \tilde{\tau}_{\br}^x \tilde{\tau}_{\br'}^x 
     + 	\tilde{J}_{y} \tilde{\tau}_{\br}^y \tilde{\tau}_{\br'}^y
     + \tilde{J}_{z} \tilde{\tau}_{\br}^z \tilde{\tau}_{\br'}^z \text{,}
\end{equation}
and discuss its dQSI and oQSI phases.

First, we note that $H_{\text{XYZ}}$ has an extra $Z_2 \times Z_2$ spin symmetry, which is not expected to be preserved upon including longer-range or multi-spin exchange.  Keeping this in mind, for simplicity we have confined our attention to the nearest-neighbor model.

The XYZ Hamiltonian has no quantum Monte Carlo sign problem over a substantial portion of its parameter space.  This is seen upon expressing $H_{\text{XYZ}}$ in terms of $\tilde{\tau}^z$ and $\tilde{\tau}^{\pm} = \tilde{\tau}^x \pm i \tilde{\tau}^y$, where
\begin{eqnarray}
H_{\text{XYZ}} &=& \sum_{\langle {\boldsymbol{r}} \boldsymbol{r}' \rangle} \Big[ {J}_{zz} \tilde{\tau}_{\boldsymbol{r}}^z \tilde{\tau}_{\boldsymbol{r}'}^z  - {J}_{\pm} (\tilde{\tau}_{\boldsymbol{r}}^+ \tilde{\tau}_{\boldsymbol{r}'}^- + h.c.) 
\nonumber \\
&& + J_{\pm\pm} ( \tilde{\tau}_{\boldsymbol{r}}^+ \tilde{\tau}_{\boldsymbol{r}'}^+ + h.c.  ) \Big] \text{.}
\label{eq:seq10}
\end{eqnarray} 
Here, $J_{zz} = \tilde{J}_z$, $J_{\pm} = -\frac{ 1}{4}(\tilde{J}_x + \tilde{J}_y )$, and  $J_{\pm\pm} = \frac{1}{4}( \tilde{J}_x - \tilde{J}_y )$.  
The XYZ Hamiltonian is thus similar to the model 
discussed in Ref.~\onlinecite{Lee2012},  but is simpler in that it lacks the bond-dependent phase factors of the latter model.  Because the transformation $\tilde{\tau}^+ \to i \tilde{\tau}^+$ sends $J_{\pm \pm} \to - J_{\pm \pm}$ without affecting the other terms, there will be no sign problem when $\tilde{J}_x + \tilde{J}_y <0$, in world-line or stochastic series expansion quantum Monte Carlo.  By cubic permutations, the sign problem is also absent when  $\tilde{J}_y+\tilde{J}_z<0$ or $\tilde{J}_x + \tilde{J}_z < 0$.

Now,  we discuss quantum spin ice in the perturbative regime \cite{Hermele2004}, where $J_{zz}  > 0$ and $J_{zz} \gg |J_{\pm}|, |J_{\pm\pm}|$. In the limit of $J_{\pm} \rightarrow 0$ and $J_{\pm\pm} \rightarrow 0$, 
the resulting Hamiltonian produces an extensively degenerate ground state 
manifold that is spanned by the so-called ``two-in-two-out'' spin ice configurations. 

Quantum dynamics is turned on with small $J_{\pm}$ 
and $J_{\pm\pm}$.
Standard degenerate perturbation theory generates an effective 
low energy Hamiltonian that acts within the spin-ice manifold.  The leading-order effective Hamiltonian is
\begin{equation}
H_{\text{eff}} = J_{\text{ring}} \sum_{\text{hexagon}} \big( 
\tilde{\tau}^+_1 \tilde{\tau}^-_2 \tilde{\tau}^+_3 \tilde{\tau}_4^- \tilde{\tau}_5^+ \tilde{\tau}^-_6 
+ h.c.,
\big)
\label{eq:effham}
\end{equation}
where $1,2,\cdots ,6$ label the 6 spins on the perimeter of a pyrochlore hexagon, and $J_{\text{ring}} \propto J_{\pm}^3 / J_{zz}^2$.  $J_{\pm \pm}$ does not contribute to $H_{{\rm eff}}$ at third-order. 

$H_{\text{eff}}$ can be mapped to a U(1) lattice gauge theory \cite{Hermele2004} by writing
\begin{eqnarray}
\tilde{\tau}^z_{\boldsymbol{r}} &=& E_{\bmr\bmr'}  ,  \label{eqn:efield}
\\
\tilde{\tau}^{\pm}_{\boldsymbol{r}} &=& e^{\pm i A_{\bmr\bmr'}}, \label{eqn:afield}
\end{eqnarray}
where the pyrochlore site $\bsr$ corresponds to the link $\bmr \bmr'$ of the dual diamond lattice, 
and $E$ and $A$ are a lattice electric field and vector potential defined on the diamond links.  This definition holds for $\bmr$ in the diamond A-sublattice and $\bmr'$ in the diamond B-sublattice.   In order to interpret $E$ and $A$ as lattice vector fields, we choose $E_{\bmr \bmr'} = - E_{\bmr' \bmr}$ and $A_{\bmr \bmr'} = - A_{\bmr' \bmr}$.

The Hamiltonian $H_{\text{eff}}$ can be interpreted as the Maxwell term suppressing magnetic flux through each hexagon.  We focus on $J_{\pm} \geq 0$, where $J_{\text{ring}} < 0$, which favors a ground state with zero flux through each hexagon.

Quantum spin ice can be understood as the deconfined phase of ${\rm U}(1)$ gauge theory \cite{Hermele2004}, and is in fact the ground state of the ring exchange Hamiltonian $H_{\text{eff}}$ \cite{Banerjee2008}.  The low-energy effective Hamiltonian density is simply ${\cal H} = K_e \vec{E}^2 + K_b \vec{B}^2$, where $\vec{E}$, $\vec{B}$ are continuum electric and magnetic fields.  So far we have been describing dQSI.   Upon permuting the axes in pseudospin space, the same discussion applies to oQSI, which arises when $\tilde{J}_y > 0$ is large.  The difference lies only in the space group transformations of the electric and magnetic fields.

In the continuum theory, we consider transformations of $\vec{E}$ and $\vec{B}$ under the $O_h$ point group to distinguish dQSI and oQSI.  
It is straightforward to identify the irreducible representations of $O_h$ under which $\vec{E}$ and $\vec{B}$ transform.  In dQSI, $\vec{E}$ transforms as $\Gamma^+_4$ (pseudovector representation) with $\vec{B}$ transforming as $\Gamma^-_4$ (vector representation).  In oQSI, on the other hand,  $\vec{E}$ transforms as $\Gamma^+_5$, with $\vec{B}$ transforming as $\Gamma^-_5$.  In both dQSI and oQSI, $\vec{E}$ ($\vec{B}$) is odd (even) under time reversal.  

The oQSI transformations can be simply understood by noting that $O_h = T_d \times Z_2$, where $T_d$ is the group of symmetries of a pyrochlore tetrahedron, and the $Z_2$ is generated by inversion ${\cal I}$.  Then any $g \in O_h$ can be uniquely written $g = {\cal I}^s t$, where $s = 0,1$, and $t \in T_d$.  Letting $D_{\Gamma^{\pm}_{4,5}}(g)$ be the representation matrices for $g \in O_h$, we have
\begin{eqnarray}
D_{\Gamma^{\pm}_5} ( t ) &=& D_{\Gamma^{\mp}_4 } ( t) \\
D_{\Gamma^{\pm}_5} ( {\cal I} t) &=& - D_{\Gamma^{\mp}_4} ({\cal I} t )\text{.}
\end{eqnarray}
Therefore, we can say that $\Gamma^+_5$ agrees with the vector representation on $T_d$, but for $g \notin T_d$, the transformations come with an extra minus sign.  Similarly, $\Gamma^-_5$ agrees with the pseudovector representation on $T_d$.

In dQSI, equal-time dipolar spin correlations are given by $\langle \vec{E} \vec{E} \rangle$ electric field correlations, which fall off as $1/r^4$.  The above results can be used to determine the corresponding (but more subtle) result for oQSI.  First, we note that $\tau^z_{\br}$ can be viewed as a vector field on the diamond lattice, transforming as a time-reversal odd pseudovector (\emph{i.e.} identical to $\vec{E}$ in dQSI).  Therefore, in the long wavelength limit, $\tau^z_{\br}$ transforms as $\Gamma^+_4$.  

To proceed, we need to construct the operator in the (Gaussian) oQSI continuum theory with smallest scaling dimension, that also transforms as $\Gamma^+_4$ and is time-reversal odd.  We have $\operatorname{dim} \vec{E} = \operatorname{dim} \vec{B} = 2$, and $\operatorname{dim} \partial_\mu = 1$.  Also, the derivative $\partial_{\mu}$ transforms as $\Gamma^-_4$.  For example, we need to consider operators of the form $\partial_{\mu} {\vec E}_\nu$, which transforms as $\Gamma^-_4 \otimes \Gamma^+_5$.  Decomposing this into irreducible representations, we find that $\Gamma^+_4$ does not appear in the tensor product, and this operator does not contribute to the dipolar spin correlations.  Proceeding in this fashion, the desired operator is instead of the form ${\cal O}_{\mu \nu \lambda} = \partial_{\mu} \partial_{\nu} (\vec{E})_\lambda$, with $\operatorname{dim} {\cal O}_{\mu \nu \lambda} = 4$.  The corresponding correlations fall off as a power law with exponent twice the scaling dimension, so the oQSI  dipolar correlations fall off as $1/r^8$.  

This result ignores the role of long-range dipolar interaction, which is potentially significant in $f$-electron systems, but its main purpose is to illustrate a sharp difference between dQSI and oQSI.  In addition, if one restricts to the XYZ Hamiltonian only (\emph{i.e.} includes no longer-range exchange), the $Z_2 \times Z_2$ symmetry actually implies that dipolar correlations fall off exponentially in oQSI, since both $\tau^z$ and $\tau^x$ transform non-trivially under $Z_2 \times Z_2$.

\section{Gauge Mean Field Theory}

The formalism of gauge mean field theory (gMFT) for the pyrochlore lattice was introduced in Refs.~\onlinecite{Savary2012,Lee2012}.  This mean-field theory is anchored to the QSI phase known to occur in the easy-axis limit \cite{Banerjee2008}, and allows one to assess the competition between QSI and  magnetically ordered phases.  Here, we adapt the gMFT formalism specifically to the pyrochlore XYZ model.

\subsection{Slave particles} 

The ground state of $H_{{\rm eff}}$ [Eq.~(\ref{eq:effham})] is a U(1) quantum spin liquid whose low
energy physics is described by compact quantum electrodynamics in $3+1$
dimensions\cite{Hermele2004,Banerjee2008}.
In the gauge theory language, the ``two-in-two-out'' spin ice rule becomes Gauss' law, and the $\tilde{\tau}^{\pm}_{\boldsymbol{r}}$ breaks the ice rule by creating electrically charged spinon
excitations on neighboring tetrahedra. 
The $J_{\pm}$ term describes the hopping of  spinons on the dual diamond lattice sites. 

Following Refs.~\onlinecite{Savary2012,Lee2012}, to make the spinons and gauge field explicit, we enlarge the physical Hilbert space by writing the spin operators as
\begin{eqnarray} 
\tilde{\tau}^{+}_{\boldsymbol{\mathsf{r}}, \boldsymbol{\mathsf{r}} + {\bf e}_{i}} &=& \Phi_{ \boldsymbol{\mathsf{r}}}^{\dagger} s^{+}_{\boldsymbol{\mathsf{r}},\boldsymbol{\mathsf{r}}+{\bf e}_i} \Phi_{ \boldsymbol{\mathsf{r}}+{\bf e}_i}^{\phantom\dagger} \\
\tilde{\tau}^z_{ \boldsymbol{\mathsf{r}}, \boldsymbol{\mathsf{r}} + {\bf e}_{i}} &=& 
s^z_{ \boldsymbol{\mathsf{r}}, \boldsymbol{ \mathsf{r} }+{\bf e}_{i}},
\end{eqnarray}
where $\boldsymbol{\mathsf{r}}$ is an A sublattice site of the diamond lattice, and
${\bf e}_{i}$ connects $\boldsymbol{\mathsf{r}}$ to its neighbors on the dual diamond 
lattice. 
$\Phi^{\dagger}_{\boldsymbol{\mathsf{r}}}$ ($\Phi^{\phantom\dagger}_{\boldsymbol{\mathsf{r}}}$) is the spinon 
creation (annihilation) operator at site ${\boldsymbol{\mathsf{r}}}$, and 
$s^z_{\boldsymbol{\mathsf{r}} \boldsymbol{\mathsf{r}}'}, s^{\pm}_{\boldsymbol{\mathsf{r}} \boldsymbol{\mathsf{r}}'}$ are spin-$1/2$ operators that act as gauge fields. 
Since the spinons are bosonic, we further write $\Phi^{\dagger}_{\boldsymbol{\mathsf{r}}} = 
e^{ i \phi_{ \boldsymbol{\mathsf{r}}  }  }$ ($\Phi^{\phantom\dagger}_{\boldsymbol{\mathsf{r}}}
= e^{ - i \phi_{ \boldsymbol{\mathsf{r}}  }  }$),
where $\phi_{ \boldsymbol{\mathsf{r}}  }  $ is a $2\pi$ periodic angular variable
and $\Phi^{\dagger}_{\bmr} \Phi^{\phantom\dagger}_{\bmr} =1$ by construction.   
In the above equations, the physical Hilbert space has been enlarged to the 
the combined space of the spinons and gauge field. 
To project back to the physical Hilbert space, we implement the following constraint,
\begin{equation}
Q_{\boldsymbol{\mathsf{r}}} = \eta_{ \boldsymbol{\mathsf{r}} } 
\sum_i s^z_{  \boldsymbol{\mathsf{r}}, \boldsymbol{\mathsf{r}} + \eta_{\boldsymbol{\mathsf{r}}} {\bf e}_i},
\end{equation}
where $\eta_{ \boldsymbol{\mathsf{r}}  } = \pm 1$ for $\boldsymbol{\mathsf{r}}   \in $ A/B sublattice. Here $Q_{\boldsymbol{\mathsf{r}}}$ is the spinon number operator and satisfies 
\begin{equation}
[\phi_{\boldsymbol{\mathsf{r}} } , Q_{ \boldsymbol{\mathsf{r}}'  }  ] = i \delta_{\boldsymbol{\mathsf{r}}\boldsymbol{\mathsf{r}}' }. 
\end{equation}

\begin{widetext}

The XYZ model Hamiltonian [Eq.~\eqref{eq:seq10}] can  be rewritten as
\begin{eqnarray}
H_{\text{XYZ}} &=& \frac{J_{zz}  }{2}  \sum_{ \boldsymbol{ \mathsf{r} }  }  Q_{\bmr}^2
- J_{\pm} \sum_{{\bmr}}\sum_{i\neq j} 
\Phi^{\dagger}_{ \bmr + \eta_{\bmr} {\bf e}_i } \Phi^{\phantom\dagger}_{\bmr + \eta_{\bmr} {\bf e}_j }
s^{-\eta_{\bmr}}_{\bmr,\bmr+\eta_{\bmr} {\bf e}_i } s^{ + \eta_{\bmr} }_{ \bmr, \bmr + \eta_{\bmr} {\bf e}_j  } 
\nonumber
 \\
&& + \frac{J_{\pm\pm} }{2} \sum_{ {\bmr}}\sum_{i\neq j} 
\big(
\Phi^{\dagger}_{\bmr}\Phi^{\dagger}_{\bmr}\Phi_{{\bmr}+\eta_{\bmr} {\bf e}_i}^{\phantom\dagger}
\Phi_{{\bmr} + \eta_{\bmr} {\bf e}_j }^{\phantom\dagger} 
s^{\eta_{\bmr}}_{{\bmr},{\bmr} + \eta_{\bmr} {\bf e}_i } 
s^{\eta_{\bmr}}_{ {\bmr}, {\bmr} + \eta_{\bmr} {\bf e}_j  }
 + {h.c.} 
\big) 
\nonumber \\
&& +  {\text{ constant.}}
\label{eq:seq15}
\end{eqnarray}
The $J_{\pm\pm}$ term now appears as an interaction between spinons. 
The above Hamiltonian is manifestly invariant under the local U(1) gauge transformation ($\Phi_{\bmr} \rightarrow \Phi_{\bmr} e^{-i \chi_{\bmr}},
s^{\pm}_{\bmr \bmr'} \rightarrow s^{\pm}_{\bmr \bmr'} e^{\pm i (\chi_{\bmr'} -\chi_{\bmr}) } $).

\subsection{Mean field theory}

Following Ref.~\onlinecite{Lee2012}, we now decouple the Hamiltonian in Eq.~\eqref{eq:seq15}
by mean field theory. As an illustration, the spinon hopping term is decoupled as follows,
\begin{eqnarray}
\Phi^{\dagger}_{ \bmr + \eta_{\bmr} {\bf e}_i } \Phi^{\phantom\dagger}_{\bmr + \eta_{\bmr} {\bf e}_j }
s^{-\eta_{\bmr}}_{\bmr,\bmr+\eta_{\bmr} {\bf e}_i } s^{ + \eta_{\bmr} }_{ \bmr, \bmr + \eta_{\bmr} {\bf e}_j  } &\rightarrow&  
\big( 
\Phi^{\dagger}_{\bmr + \eta_{\bmr} {\bf e}_i} \Phi^{\phantom\dagger}_{\bmr + \eta_{\bmr}{\bf e}_j } - \langle \Phi^{\dagger}_{\bmr + \eta_{\bmr} {\bf e}_i} \Phi_{\bmr + \eta_{\bmr} {\bf e}_j}^{\phantom\dagger} \rangle
\big)
\langle s^{-\eta_{\bmr}}_{\bmr,\bmr+\eta_{\bmr} {\bf e}_i }  \rangle
\langle s^{+\eta_{\bmr}}_{\bmr,\bmr+\eta_{\bmr} {\bf e}_j } \rangle
\nonumber \\
&+& 
\langle 
\Phi^{\dagger}_{ \bmr + \eta_{\bmr} {\bf e}_i } \Phi^{\phantom\dagger}_{\bmr + \eta_{\bmr} {\bf e}_j }
\rangle
\big(
s^{-\eta_{\bmr}}_{\bmr,\bmr+\eta_{\bmr} {\bf e}_i } \langle s^{+\eta_{\bmr}}_{\bmr,\bmr + \eta_{\bmr} {\bf e}_j }  \rangle 
+ \langle s^{-\eta_{\bmr}}_{\bmr,\bmr+\eta_{\bmr} {\bf e}_i } \rangle 
s^{+\eta_{\bmr}}_{\bmr,\bmr + \eta_{\bmr} {\bf e}_j }  
- \langle s^{-\eta_{\bmr}}_{\bmr,\bmr+\eta_{\bmr} {\bf e}_i } \rangle 
   \langle s^{+\eta_{\bmr}}_{\bmr,\bmr + \eta_{\bmr} {\bf e}_j }  \rangle
\big).
\end{eqnarray}
Similar decouplings can also be made to the $J_{\pm\pm}$ term. 
The microscopic Hamiltonian is now reduced to 
mean field Hamiltonians for both spinon sector $H_{\Phi}$ and gauge sector $H_s$,
\begin{equation}
H_{\text{XYZ}} \rightarrow H_{\text{gMFT}} = H_{\Phi} + H_s.
\end{equation}
Here, $H_{\Phi}$ is given by
\begin{eqnarray}
H_{\Phi} &=& \frac{J_{zz}}{2} \sum_{\bmr} Q_{\bmr}^2
 - J_{\pm} \Delta^2 \sum_{\bmr} \Phi^{\dagger}_{ {\bmr} + \eta_{\bmr} {\bf e}_i  } 
\Phi^{\phantom\dagger}_{{\bmr} + \eta_{\bmr} {\bf e}_j}
\nonumber \\ 
&& + \frac{J_{\pm\pm} \Delta^2 }{2} \sum_{ \bmr \in \text{A} } \sum_{i \neq j}
\Big[
{\Phi^{\dagger}_{\bmr}} \Phi^{\dagger}_{\bmr}   \chi^{\text{B}}_{ij} 
+ {\chi_0^{\text{A}} }^{\ast} \Phi_{ \bmr + {\bf e}_i }^{\phantom\dagger} \Phi_{ \bmr +{\bf e}_j }^{\phantom\dagger}
+ 2 \big( \Phi_{\bmr}^{\dagger} \Phi_{\bmr + {\bf e}_i}^{\phantom\dagger} \xi_j   + 
 \Phi_{\bmr}^{\dagger} \Phi_{\bmr + {\bf e}_j}^{\phantom\dagger} \xi_i \big)
+ h.c. 
\Big] 
\nonumber \\
&& + \frac{J_{\pm\pm} \Delta^2 }{2} \sum_{ \bmr \in \text{B} } 
\sum_{i \neq j}
\Big[
{\Phi^{\dagger}_{\bmr}} \Phi^{\dagger}_{\bmr}   \chi^{\text{A}}_{ij} 
+ {\chi_0^{\text{B}} }^{\ast} \Phi_{ \bmr - {\bf e}_i }^{\phantom\dagger} \Phi_{ \bmr - {\bf e}_j }^{\phantom\dagger}
+ 2 \big( \Phi_{\bmr}^{\dagger} \Phi_{\bmr - {\bf e}_i}^{\phantom\dagger} \xi_j^{\ast}   + 
 \Phi_{\bmr}^{\dagger} \Phi_{\bmr - {\bf e}_j}^{\phantom\dagger} \xi_i^{\ast} \big)
+ h.c. 
\Big].
\label{eq:hamphi}
\end{eqnarray}
\end{widetext}
We have chosen a mean-field state where the spinons feel zero magnetic flux through each hexagon, as appropriate for $J_{{\rm ring}} < 0$ and $J_{\pm} < 0$, and have thus chosen a gauge in which the spinon hopping is uniform.
We have introduced the sublattice mixing parameters,
\begin{equation}
\xi_i \equiv \langle   \Phi^{\dagger}_{\bmr} \Phi^{\phantom\dagger}_{\bmr + \eta_{\bmr} {\bf e}_i}   \rangle
\quad \text{for} \quad \bmr \in \text{A},
\end{equation}
the onsite pairing parameters,
\begin{eqnarray}
\chi_0^{\text{A}} & = & \langle  \Phi_{\bmr}^{\phantom\dagger} \Phi_{\bmr}^{\phantom\dagger}  \rangle
\quad \text{for} \quad \bmr \in \text{A}, 
\\
\chi_0^{\text{B}} & = & \langle  \Phi_{\bmr}^{\phantom\dagger} \Phi_{\bmr}^{\phantom\dagger}  \rangle
\quad \text{for} \quad \bmr \in \text{B}, 
\end{eqnarray}
and the inter-site pairing 
\begin{eqnarray}
\chi^{\text{A}}_{ij} &=& 
\langle \Phi^{\phantom\dagger}_{\bmr - {\bf e}_i}\Phi^{\phantom\dagger}_{\bmr - {\bf e}_j} \rangle
\quad \text{for} \quad \bmr \in \text{B},
\\
\chi^{\text{B}}_{ij} &=& 
\langle \Phi^{\phantom\dagger}_{\bmr + {\bf e}_i}\Phi^{\phantom\dagger}_{\bmr + {\bf e}_j} \rangle
\quad \text{for} \quad \bmr \in \text{A}.
\end{eqnarray}
Finally, the parameter $\Delta$ is defined as $\Delta = \langle s^{\pm}_{\bmr,\bmr \pm {\bf e}_i} \rangle$
is chosen uniformly on all bonds due to the above gauge choice.
$H_{s}$ only contains Zeeman terms for $s^{\mu}_{\bmr \bmr'}$ and is trivially solved, 
leading to $\Delta = 1/2$.

The ground state of $H_{\Phi}$ in Eq.~\eqref{eq:hamphi} is then solved self-consistently
under the constraint $\Phi_{\bmr}^{\dagger} \Phi_{\bmr}^{\phantom\dagger} =1$. 
The gMFT ground state is selected by optimizing the variational energy 
$\langle H_{\text{XYZ}} \rangle$ in Eq.~\eqref{eq:seq15}. 
Magnetic ordering appears when the spinon field is condensed with the physical order
parameter given by 
$\langle \tilde{\tau}^{+}_{ \bmr,\bmr+ {\bf e}_i } \rangle = \langle \Phi^{\dagger}_{\bmr} \rangle
\langle s^{+}_{\bmr, \bmr + {\bf e}_i}  \rangle 
\langle \Phi^{\phantom\dagger}_{\bmr + {\bf e}_i} \rangle$.
We find that, the pairing parameters always vanish when the spinon field is not condensed, indicating 
the absence of an intermediate $Z_2$ quantum spin liquid in this mean field approach. Therefore,
the phase boundary of the phase diagram in the main text is obtained when the spinon field  condensation takes place.

\end{document}